\documentclass[]{IEEEtran}
\usepackage{mathrsfs}
\usepackage{amsfonts}
\usepackage{graphicx,cite,epsfig,amssymb,amsmath}
\usepackage{color,xcolor}
\definecolor{red}{rgb}{1.00, 0.00, 0.00}
\usepackage{pifont}
\usepackage{stmaryrd}
\usepackage{setspace}
\usepackage{subfigure}
\usepackage{cite}
\usepackage{array}
\usepackage{float}
\usepackage{verbatim}
\usepackage{multirow}
\usepackage{booktabs}
\usepackage{makecell}
\usepackage[linesnumbered, ruled]{algorithm2e}
\usepackage{epstopdf}
\usepackage{mathtools}
\usepackage{diagbox}
\usepackage{cases}
\usepackage{amsthm}
\usepackage[colorlinks,
linkcolor = black,
anchorcolor = black,
citecolor = black]{hyperref}
\usepackage{textcomp}

\providecommand{\algorithmname}{Algorithm}
\floatname{algorithm}{\protect\algorithmname}
\newcommand{\bm}[1]{\mbox{\boldmath{$#1$}}}
\newtheorem{thm}{Theorem}

\newtheorem{lem}{Lemma}

\usepackage{arydshln}
\allowdisplaybreaks[4]

\newtheoremstyle{noparens}%
  {}{}%
  {\itshape}{}%
  {\bfseries}{.}%
  { }%
  {\thmname{#1}\thmnumber{ #2}\mdseries\thmnote{ #3}}
\theoremstyle{noparens}

\begin{document}

\title{Task-Oriented Multimodal Edge Intelligence via Integrated Sensing-Communication-Computation}
\author{\IEEEauthorblockN{Weiwei~Chen, Yinghui~He,~\IEEEmembership{Member,~IEEE}, Zhong~Ye, Dingzhu~Wen,~\IEEEmembership{Member,~IEEE},\\Guanding~Yu,~\IEEEmembership{Senior Member,~IEEE}}
\thanks{
Manuscript received 4 December 2025; revised 22 March 2026 and 23 May 2026; accepted 4 July 2026. (\emph{Corresponding author: Zhong Ye})

W. Chen and Z. Ye are with the College of Information Science and Electronic Engineering, Zhejiang University, Hangzhou 310027, China (e-mail: \{22331145, 12431118\}@zju.edu.cn).

Y. He is with the College of Computing and Data Science, Nanyang Technological University, 639798, Singapore (email: yinghui.he@ntu.edu.sg).

G. Yu is with the State Key Laboratory of Ocean Sensing, Zhoushan 316021, China, and also with the College of Information Science and Electronic Engineering, Zhejiang University, Hangzhou 310027, China (email: yuguanding@zju.edu.cn).

D. Wen is with the School of Information Science and Technology, ShanghaiTech University, Shanghai 201210, China (e-mail: wendzh@shanghaitech.edu.cn).}}

\markboth{IEEE Transactions on Wireless Communication,~Vol.~xx, No.~xx, xx~2026}%
{Shell \MakeLowercase{\textit{et al.}}: A Sample Article Using IEEEtran.cls for IEEE Journals}

\maketitle
\begin{abstract}
Integrated sensing, communication, and computation (ISCC) has recently emerged as a unified framework for enabling edge intelligence. However, existing ISCC designs predominantly rely on single-modal sensing, which is inherently vulnerable to occlusions, environmental uncertainties, and modality-specific failures, leading to degraded robustness in real-world deployments. This motivates the need for multi-modal ISCC, yet its design remains insufficiently explored. 
Compared with the single-modal case, multi-modal ISCC is more challenging because heterogeneous modalities enlarge data dimensionality and tighten communication/computation/energy budgets, while inter-modal correlations further complicate performance characterization.
To address these challenges, we propose a task-oriented multi-modal ISCC framework that integrates device-side feature extraction with edge-side joint multi-modal inference.
A central component of our approach is the maximal coding rate reduction (MCR$^2$) criterion, which enables each device to learn compact and discriminative task-relevant features, offering clear advantages over conventional cross-entropy–based extractors. We further leverage MCR$^2$ as a principled metric for edge-side sensing evaluation. On this basis, we formulate a sensing accuracy maximization problem under delay and resource constraints and develop an efficient block coordinate descent (BCD) algorithm after transforming the problem into a more tractable equivalent form.
Focusing on a human activity recognition task, we conduct extensive experiments on publicly available datasets to evaluate the performance of the proposed ISCC framework. The results demonstrate that our approach consistently outperforms three baseline schemes under limited resource conditions.
\end{abstract}


\section{Introduction}
The upcoming sixth-generation (6G) wireless networks are expected to go beyond traditional communication services, ushering in an intelligent and seamlessly connected era powered by edge intelligence and supporting emerging applications, such as autonomous driving, smart cities, and immersive virtual environments~\cite{cui2021integrating,kaushik2024toward,du2024distributed}. The realization of these applications hinges on the tight integration of three fundamental processes: sensing to acquire environmental information, communication to convey it, and computation to support downstream intelligent decision-making.
However, the existing wireless infrastructure, primarily designed for data transmission, struggles to meet the joint demands of high-precision sensing and intelligent computation.
A key reason for this mismatch is that traditional pipelines treat sensing, communication, and computation as isolated components, preventing coordinated optimization.
Furthermore, the escalating computation demands of edge-intelligent services further intensify the burden on limited computation resources. To address these limitations, a new technique called integrated sensing, communication, and computation (ISCC) has been recently proposed~\cite{zhu2023pushing,wen2024survey}. ISCC synergistically combines integrated sensing and communication (ISAC)~\cite{isac1,isac2,isac3,11288062} with mobile edge computing (MEC)~\cite{xiao2022multi,sabella2016mobile,hevesli2024task} to enable joint resource sharing and cross-layer optimization for enhanced performance and efficiency.

Considering that intelligent sensing tasks are both computationally intensive and delay sensitive, task-oriented ISCC systems require an efficient sensing execution framework together with the joint allocation of sensing, communication, and computation resources.
Most existing task-oriented ISCC approaches~\cite{wen2023taskair,wang2023device,chen2025sensing,ding2022joint,liu2025joint,zhao2024multi,wen2023task,wang2024joint,he2023integrated} adopt such a framework but are fundamentally built upon a single sensing modality.
This single-modality assumption greatly limits system robustness, as it makes the sensing process highly vulnerable to occlusions, environmental uncertainties, and modality-specific failures, often leading to notable performance degradation in practical deployments.
To overcome these limitations, multi-modal sensing has emerged as a natural extension, as heterogeneous modalities can provide complementary and redundant information that enhances robustness and task accuracy~\cite{wang2024multi,yang2023mm}.
Nevertheless, incorporating multi-modal sensing into ISCC systems introduces the following three challenges.
\begin{itemize}
    \item  \textit{Increased data volume:} Incorporating multiple sensing modalities substantially increases the volume of sensing data that must be acquired, processed, and transmitted. The resulting high-dimensional data streams impose significant pressure on wireless communication links and may lead to prohibitive delays, especially in resource-constrained environments.

    \item \textit{Performance evaluation metric:} 
    Intelligent sensing tasks typically rely on deep learning–based artificial intelligence (AI) models for inference, whose inherent ``black-box'' nature makes it challenging to derive tractable analytical metrics that accurately reflect task performance. This lack of transparent and differentiable accuracy metrics complicates system-level optimization.
    
    \item \textit{Joint resource allocation:} The sensing, communication, and computation processes across different modalities are tightly coupled and must share limited communication bandwidth, energy, and computational resources. These interdependencies, combined with the dual requirements of low latency and high accuracy in intelligent sensing tasks, make the joint resource allocation considerably more complex.
    
\end{itemize}


To overcome the above challenges, we develop a task-oriented multi-modal ISCC framework that integrates multi-modal feature extraction with joint sensing, communication, and computation resource allocation.
In multi-modal systems, sensing signals generated by different modalities often differ significantly in dimensionality, statistical properties, and task relevance. Therefore, instead of transmitting raw modality-specific signals to the edge server, each device first performs local feature extraction to obtain a compact and task-aligned representation that facilitates efficient downstream processing.
In particular, we adopt the maximal coding rate reduction (MCR$^2$) criterion proposed in~\cite{yu2020learning} as the feature extraction strategy, enabling each device to learn discriminative yet compact feature representations.
Furthermore, we leverage the MCR$^2$ objective as an alternative sensing performance metric at the edge server, allowing us to evaluate the sensing quality in a principled and differentiable manner. Based on this metric, we develop a block coordinate descent (BCD)-based optimization algorithm to jointly allocate sensing, communication, and computation resources to maximize the overall sensing performance.
The main contributions of this paper are summarized as follows.
\begin{itemize}
    \item We propose a task-oriented multi-modal ISCC framework that integrates local feature extraction at devices with joint multi-modal recognition at the edge server. Focusing on a human activity recognition task, we employ the MCR$^2$ criterion to guide the feature extractor in learning compact, discriminative, and task-relevant representations.
    
    \item We provide a quantitative performance analysis of the proposed framework, revealing how quantization distortion, transmit power, and communication capacity jointly influence sensing performance. By adopting MCR$^2$ as a principled metric for sensing accuracy, we formulate a sensing accuracy maximization problem under delay, energy, and resource constraints.

    \item To solve the resulting problem, we develop an efficient BCD-based optimization algorithm. We first transform the original problem into an equivalent but more tractable form, and then iteratively solve it via block coordinate descent, where each iteration solves a convex subproblem to update quantization bit allocation and communication time allocation.

    \item We conduct extensive experiments on publicly available datasets to evaluate the proposed framework. Both support vector machines (SVM) and multilayer perceptrons (MLP) are adopted as edge classifiers. Experiment results validate the effectiveness of using MCR$^2$ as a sensing metric and show that the proposed multi-modal ISCC scheme significantly outperforms all baselines, demonstrating the clear advantage of multi-modal sensing and joint resource optimization.
\end{itemize}

The remainder of this paper is organized as follows. Section~\ref{sec:related} reviews the related work. Section~\ref{sec:sys_model} introduces the multi-modal ISCC system, and elaborates on the sensing model, feature extractor, communication model, and edge computing model. Section~\ref{sec:reformulation} formulates the sensing accuracy maximization problem and transforms it into a more tractable form. Then, a BCD-based algorithm is proposed in Section~\ref{sec:BCD} to solve the problem. Section~\ref{sec:test_all} presents the experiment results for verifying the performance of the proposal, and Section~\ref{sec:conclu} concludes the paper.

\textit{Notations}: In this paper, scalars are denoted by non-bold letters, vectors are denoted by boldface lowercase letters, and matrices are denoted by boldface uppercase letters.
$\mathbb{R}$ and $\mathbb{C}$ represent the sets of real and complex numbers, respectively.
For a matrix $\bm{A}$, 
 $\mathrm{tr}(\bm{A})$ denotes its trace, $\mathrm{diag}(\bm{A})$ denotes a vector formed by the diagonal entries of 
 $\bm{A}$,
$\bm{A}^{\mathrm{T}}$ denotes its transpose, and $\bm{A}^{-1}$ denotes its inverse. $\|\cdot\|_2$ represents the Euclidean norm.  $\mathcal{N}^{\mathrm{v}}(\bm{w};\bm{\mu},\bm{\Sigma},\bm{\Gamma})$ denotes a Gaussian distribution of vector $\bm{w}$ with mean $\bm{\mu}$, covariance $\bm{\Sigma}$, and relation matrix $\bm{\Gamma}$.  $\mathcal{N}(\mu, \sigma^2)$ represents a scalar Gaussian distribution with mean $\mu$ and variance $\sigma^2$.

\section{Related Work}\label{sec:related}
This work is most closely related to research on ISCC, which can be broadly categorized into two lines: ISCC for traditional sensing tasks (e.g., detection and tracking)~\cite{ding2022joint,liu2025joint,zhao2024multi} and ISCC for AI-driven tasks~\cite{wen2023task,wang2024joint,he2023integrated}.
For traditional sensing tasks, the authors in~\cite{ding2022joint} formulated a multi-objective optimization problem that jointly considers energy consumption and radar beampattern design, and then optimized transmit precoding together with wireless resource allocation to enhance the sensing performance.
The authors in~\cite{liu2025joint} investigated joint beamforming and offloading design within a three-tier ISCC framework. The task execution time was minimized while ensuring sening performance, as measured by the signal-to-interference-plus-noise ratio (SINR). However, SINR has limitations as a metric for tasks such as target detection and collision avoidance.
To address this, authors in~\cite{zhao2024multi} incorporated the Cramér-Rao bound (CRB) into the ISCC design, and then minimizing it under constraints related to communication, computation resources, and service requirements.
For AI-driven tasks,~\cite{wen2023task} introduced the discrimination gain metric to analytically capture the influence of sensing, communication, and computation resources on task accuracy, and further designed a task-oriented resource allocation scheme to enhance the sensing performance. Along this line,~\cite{wang2024joint} investigated an ISCC-enabled multi-view multi-task edge-AI system, where shared sensory data from multiple devices are jointly used to support several inference tasks with reduced sensing, computation, and communication overhead.
Considering that directly implementing AI algorithms to all sensing data would lead to high computation resource waste,~\cite{he2023integrated} proposed an action detection module to filter static target. Moreover, they further formulated a sensing accuracy maximization problem   with the delay requirement of computation tasks, and developed an optimal resource allocation strategy to enhance the sensing performance. Nevertheless, the above studies primarily focused on single-modal sensing, which is vulnerable to occlusions and complex environments. To address this limitation, we incorporate multi-modal sensing to enhance system robustness and task performance.

This work is also related to multi-modal sensing.
A large body of research has focused on developing various models and fusion strategies to integrate heterogeneous sensing data and improve end-to-end inference accuracy~\cite{liu2024wireless,zhang2024integrated,cheng2023intelligent,ye2025radar}.  However, such model-level approaches are inadequate for real-world deployment, where communication, energy, and computation resources are inherently limited.
To address this, recent studies have explored multi-modal semantic communication frameworks that employ neural encoders to extract ``modality'' specific semantic information.
For example, the authors in~\cite{xie2022task} proposed DeepSC-VQA, which employes a device-side semantic encoder to extract compressed task-relevant features from images and text, and an edge-side semantic decoder for multi-modal reconstruction and inference.
This design effectively reduces the uplink data volume under limited bandwidth while maintaining high task performance. Subsequently,~\cite{zhang2024unified} extended multi-modal semantic communication to multi-task scenarios and proposed a unified framework, U-DeepSC, which incorporates a vector-level dynamic feature selection module that adaptively adjusts the number of transmitted features according to task requirements and channel conditions. 
Beyond semantic communication, a further evolution is task-oriented communication, which focuses on transmitting task-essential information. 
Authors in~\cite{wan2023cooperative} proposed a cooperative strategy that exploits the low-rate transmission output of one modality to control the transmission of other modalities, thereby  balancing the task performance and communication cost. 
A robust  multi-modal task-oriented communication framework was introduced in~\cite{fu2025robust}, integrating a two-stage variational information bottleneck module with mutual information redundancy  minimization.
However, the above approaches lack a unified modeling of the tightly coupled task execution pipeline, which in turn degrades task execution efficiency and overall performance.



\section{System Model}\label{sec:sys_model}

In this section, we first present the multi-modal ISCC system, and then elaborate on the sensing model and feature extraction based on MCR$^2$, communication model, and edge computing model.

\subsection{Multi-Modal ISCC System} \label{ssec:iscc_model}

\begin{figure}[t]
	\setlength{\abovecaptionskip}{8pt} 
	\centering	
    \includegraphics[width=1\linewidth]{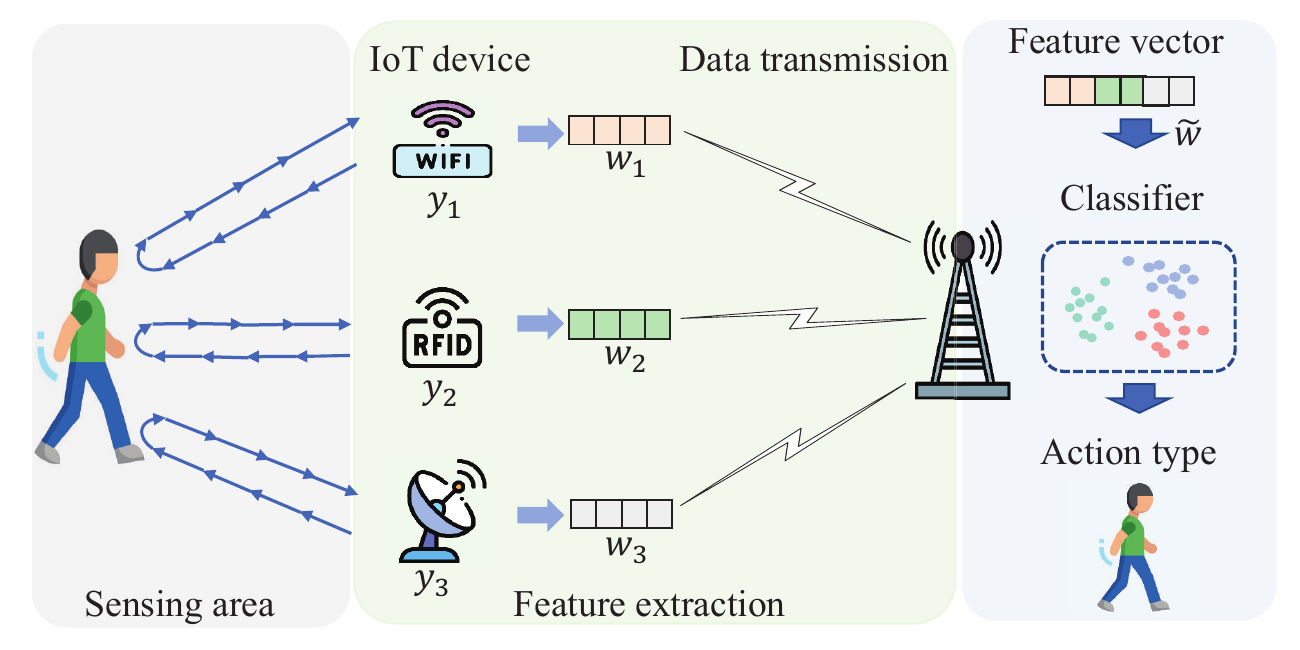}
	\vspace{-3ex}
	\caption{An ISCC system with multi-modal sensing.}
	\label{fig:sys}
    \vspace{-2ex}
\end{figure}

As illustrated in Fig.~\ref{fig:sys}, we consider a multi-modal ISCC system, consisting of 
$K$ IoT devices and
a base station (BS) equipped with an edge server.
The proposed system collects sensing signals containing both environmental and target-related information through IoT devices and transmits them to the edge server to enable edge intelligence.
Since a single sensing modality may suffer from degraded recognition accuracy under occlusions or in complex environments, multi-modal sensing is adopted to enhance robustness and reliability in this system.
Specifically, each IoT device is equipped with an individual sensing modality, such as millimeter-wave radar, RFID, or WiFi sensing.\footnote{The framework can be further extended to scenarios with a larger number of devices, where each device may adopt identical or heterogeneous modalities depending on application requirements.}  In this work, we take human activity recognition as a representative task~\cite{10556745,li2022human}, which can be applied to extended reality (XR), smart homes, intelligent healthcare, and other emerging domains.

To this end, there are four main steps.
\begin{itemize}
    \item[1)] \textit{Sensing}: Each IoT device first transmits modality-specific sensing signals toward the same target area and receives corresponding echo signals to acquire sensing data\footnote{{The heterogeneous devices are pre-configured to cover the same target area and can be coordinated via periodic triggering or time-stamping.}}. 
    \item[2)] \textit{Local processing}: Given limited communication resources, directly transmitting the raw data from each device to the edge server would cause large latency and also raise privacy concerns. To mitigate these issues, we perform local feature extraction at the device side.
    \item[3)] \textit{Data transmission}: The extracted feature vectors are then transmitted to the edge server via wireless links for further processing.
   \item[4)] \textit{Edge computing}: At the edge server, the multi-modal feature vectors are concatenated to construct a joint representation, which is subsequently fed into a classifier for action recognition.
\end{itemize}

\begin{figure}[t]
    \vspace{-1ex}
		\centering		\includegraphics[width=1\linewidth]{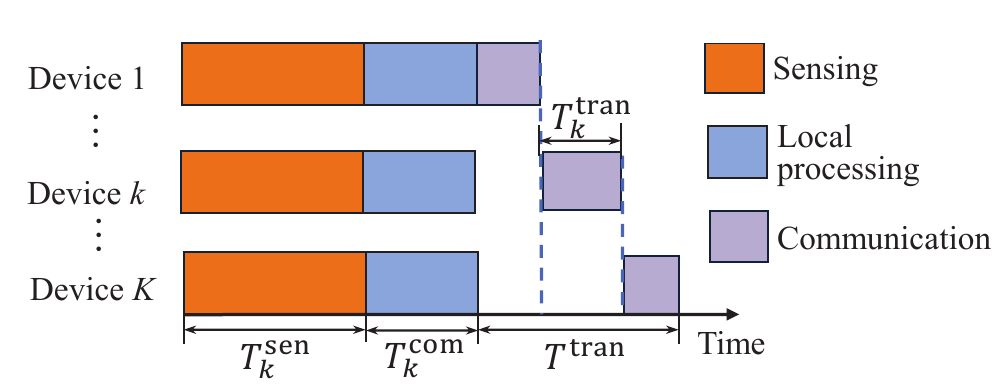}
		\vspace{-3ex}
		\caption{Delay analysis for the ISCC system. }
		\label{fig:frame}
\end{figure}

To maintain compatibility with practical IoT systems and follow the operational procedure of first sensing and then communication, a time-division structure is adopted to switch between sensing and communication, as shown in Fig.~\ref{fig:frame}. In the sensing phase, different devices operate on separate frequency bands, such that no mutual interference exists among their sensing signals. In the communication phase, all devices share the same uplink band to transmit their data to the BS. To avoid inter-device interference during uplink transmission, time-division multiple access (TDMA) is adopted for multi-device uploading\footnote{{Our proposal can be readily extended to orthogonal frequency-division multiple access (OFDMA) with appropriate modifications.}}.

For device $k$, 
its sensing time can be denoted as
$T_k^{\mathrm{sen}}$, and its local processing time is $T_k^{\mathrm{com}}$. The communication time for feature transmission is denoted as $T_k^{\mathrm{tran}}$.
Since sensing and local processing are carried out in parallel across devices, the total delay of these two steps is determined by the maximum values across all devices, i.e., $\max_k\{T_k^{\mathrm{sen}}+ T_k^{\mathrm{com}}\}$. {By contrast, feature transmission follows a TDMA protocol, so the total transmission delay is the sum of the allocated communication time over all devices, i.e., $ \sum_{k=1}^{K} T_k^{\mathrm{tran}}$, which is a controllable variable determined by the resource allocator at the BS.} Thus, the total delay can be expressed as
\begin{equation}
    \max_k\{T_k^{\mathrm{sen}}+ T_k^{\mathrm{com}}\} + \sum_{k=1}^{K} T_k^{\mathrm{tran}}.
\end{equation}
Note that the delay of the edge computation is ignored, due to the abundant computation resources at the edge server.

\subsection{Sensing Model and Feature Extraction} \label{sys:feature}

During the sensing phase, IoT device $k$ transmits a probing
signal with power $p^{\mathrm{s}}_k$
corresponding to its specific modality, and simultaneously receives the echo signals that carry environmental and target information. By comparing the transmitted and received signals, the device obtains the final sensing signal $\bm{y}_k \in \mathbb{C}^{m_k}$.
For instance, in the WiFi modality, the sensing signal can be obtained by dividing the received signal of each subcarrier by its corresponding transmitted signal, similar to the extraction of channel state information (CSI).
Transmitting all raw sensing signals directly to the edge server would lead to significant communication delays and is often unnecessary, as data from different modalities may exhibit redundancy and cross-modal correlations. Therefore, we first perform feature extraction on the sensing signals to reduce its dimensionality and transmission overhead.
In classification tasks, the feature extractor is commonly trained under the supervision of the cross-entropy loss. 
{Despite its effectiveness and widespread use, this end-to-end data fitting approach 
obscures the intrinsic geometric structures and statistical properties of the learned features, leading to reduced interpretability and robustness in downstream tasks~\cite{yu2020learning,chan2022redunet,chu2024adversarial}.
Moreover, its behavior is heavily coupled with the underlying network parameters, which would result in a highly complicated optimization problem if directly used for system-level design, thereby making it difficult to support resource allocation and adaptive multi-modal transmission policy optimization.
}


To address these issues, we adopt the  MCR$^2$ criterion as our feature extraction objective. 
This criterion seeks to learn discriminative yet compact representations by maximizing the difference in coding efficiency between intra-class and inter-class features. Specifically, features from different classes are encouraged to occupy distinct subspaces that collectively span a space with the maximum possible volume, where the coding rate quantitatively measures this volume or information content.
Conversely, features belonging to the same class are constrained to lie within a low-dimensional subspace of small volume, thereby enhancing intra-class compactness.
The goal of MCR$^2$ is thus to maximize the coding rate reduction, i.e., the difference between the coding rate required to jointly encode all features and that required when encoding features of each class separately.

To derive the coding rate reduction, we first analyze the coding efficiency (i.e., coding rate) for inter-class features.
Let $\bm{w}_k=[w_k[1],...,w_k[d_k]] \in  \mathbb{R}^{d_k}$ denote the feature vector after extraction at device $k$, where $d_k$ represents its feature length with $d_k\ll m_k$.
The concatenated feature vector of  all $K$ devices at the edge server is denoted as $\bm{w} \triangleq \begin{bmatrix} \bm{w}_1^\top, \dots, \bm{w}_K^\top \end{bmatrix}^\top \in \mathbb{R}^D$, where $D=\sum_{k \in \mathcal{K}}d_k$. 
Suppose there are $S$  
samples with their feature matrix being
$\bm{W} = \begin{bmatrix} \bm{w}^{(1)}, \dots, \bm{w}^{(S)} \end{bmatrix} \in \mathbb{R}^{D \times S}$, the coding rate required to represent these samples is defined by
\begin{equation} \label{eq:R_W}
R(\bm{W}, \epsilon) = \frac{1}{2} \log \det \left( \bm{I} + \frac{D}{S \epsilon^2} \bm{W} \bm{W}^\top \right),
\end{equation}
where $\epsilon$ is the distortion level.
This formula can be obtained by considering the packing of $\epsilon$-balls within the subspace spanned by $\bm{W}$ under a Gaussian source model~\cite{yu2020learning}. 
Then, we analyze the coding rate for intra-class features, i.e., the coding efficiency achieved when encoding the features of each class separately.
Assume a classification task with $L$ distinct classes, denoted by $\mathcal{L} \triangleq \{1, \dots, L\}$. 
Let $\bm{\Pi} = \left\{ \bm{\Pi}_l \in \mathbb{R}^{S \times S} \right\}_{l \in \mathcal{L}}$ represent a set of diagonal matrices, where the $s$-th diagonal entry of $\bm{\Pi}_l$ means the probability that the $s$-th sample belongs to the $l$-th class.  
Here, $s \in \mathcal{S} \triangleq \{1, \dots, S\}$ and $l \in \mathcal{L}$ are the sample and class indices, respectively, and the matrices satisfy $\sum_{l \in \mathcal{L}} \bm{\Pi}_l = \bm{I}$.
Then, the coding rate for encoding different classes of features independently up to the distortion $\epsilon$ is 
\begin{equation} \label{eq:Rc_W}
R^{\mathrm{c}}(\bm{W}, \epsilon\!\! \mid \!\!\bm{\Pi})\!\! =\!\! \sum_{l \in \mathcal{L}} \!\frac{\mathrm{tr}(\bm{\Pi}_l)}{2S} \log \!\det \!\left( \!\bm{I} \!+ \!\frac{D}{\mathrm{tr}(\bm{\Pi}_l)\, \epsilon^2} \bm{W} \bm{\Pi}_l \bm{W}^\top \!\!\right).
\end{equation}
Consequently, the coding rate reduction can be expressed as
\begin{equation}\label{equ:mcr2}
    \Delta R(\bm{W}, \bm{\Pi}, \epsilon) = R(\bm{W}, \epsilon) - R^{\mathrm{c}}(\bm{W}, \epsilon \mid \bm{\Pi}).
\end{equation}

\begin{figure}[t]
	\setlength{\abovecaptionskip}{8pt} 
		\centering		\includegraphics[width=1\linewidth]{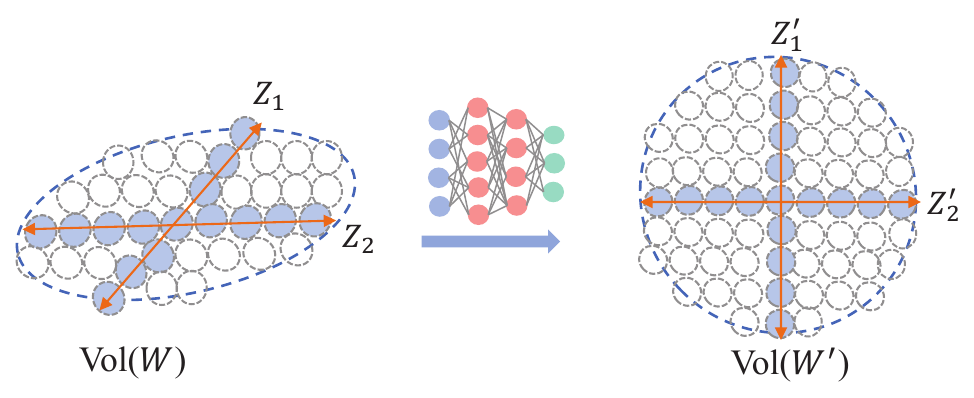}
		\vspace{-3.5ex}
		\caption{An illustration of the coding rate reduction function. Specifically, $W$ and $W^{'}$ are two representations by reduced rates, where $Z_1$ and $Z_2$ represent the subspaces corresponding to the first and second classes, respectively. Our objective is to train the neural network such that the extracted features exhibit the desirable structure illustrated on the right, i.e., maximizing the number of white balls ($\Delta R$).}
		\label{fig:mcr2}
        \vspace{-3ex}
\end{figure}

To provide an intuitive understanding of~(\ref{equ:mcr2}), Fig.~\ref{fig:mcr2} illustrates its geometric interpretation from the perspective of sphere packing.
Ideally, features from different classes should be uncorrelated and discriminative, thereby spanning a space of large volume, i.e., maximizing $R(\bm{W}, \epsilon)$. In contrast, features within the same class are expected to be compact, corresponding to a minimal volume, i.e., minimizing $R^c(\bm{W}, \epsilon\!\! \mid \!\!\bm{\Pi})$. 
These requirements are jointly captured by maximizing the coding rate reduction. Moreover, given that $\Delta R(\bm{W}, \bm{\Pi}, \epsilon)$ increases with the size of $\bm{W}$, each feature sample is normalized to guarantee fair comparison across different representations. 
Accordingly, the feature extraction objective is formulated as
\begin{subequations} \label{sec:obj_feature}
\begin{eqnarray}
&\max\limits_{\bm{W}} & \Delta R(\bm{W}, \bm{\Pi}, \epsilon), \\
&\text{s.t.}       & \left\| \bm{w}^{(s)} \right\|_2^2 = 1,\quad s \in \mathcal{S}. 
\end{eqnarray}
\end{subequations}

It is worth noting that the calculation of the coding rate involves multiple feature samples and the concatenation of features from different devices. Therefore, the feature extraction networks are firstly centrally trained on the dataset with the $\mathrm{MCR}^2$ objective, and the trained models are subsequently deployed on the local IoT devices for inference.  During inference, each device senses the common target area, performs local feature extraction independently, and then transmits its feature vector $\bm{w}_k$ to the edge server, where multi-modal features are aggregated for the final recognition task.

\subsection{Communication Model}

To enable low-cost and reliable transmission, digital communication is employed to transmit the feature vectors. Prior to transmission, feature elements are quantized at the IoT devices. Specifically, we adopt a high quantization bit range scheme~\cite{shlezinger2021deep}. For the $i$-th feature element in feature vector $\bm{w}_k$, the quantized data can be denoted as 
\begin{equation}
    z_k[i]=\sqrt{Q_{k,i}}w_k[i]+n_k, \label{eq:quanti}
\end{equation}
where $Q_{k,i}$ denotes the quantization gain, and $n_k$ represents the quantization noise modeled as a Gaussian random variable, i.e., $n_k \sim \mathcal{N}(0, \delta_k^2)$, with variance $\delta_k^2$. At the BS, the $i$-th feature element is recovered as
\begin{equation}
    \tilde{w}_k(i) = \frac{z_k[i]}{\sqrt{Q_{k,i}}} 
    = w_k[i] + \frac{n_k}{\sqrt{Q_{k,i}}}.
\end{equation}

{Under the assumption of additive Gaussian distortion, the mutual information (that characterizes the data size of the feature vector) 
between the extracted feature vector $\bm{w}_k$ and the recovered feature vector
$\tilde{\bm{w}}_k = [\tilde{w}_k[1], \dots, \tilde{w}_k[d_k]] \in \mathbb{R}^{d_k}$ can be expressed as
\begin{equation}
    I(\tilde{\bm{w}}_k; \bm{w}_k)=\sum_{i=1}^{d_k} \log_2 \left(1 + \frac{Q_{k,i}}{\delta_k^2} \right),
\end{equation}
which also represents the communication overhead of transmitting the feature vector to the edge server for device $k$.}

The overall communication bandwidth is denoted by $B$, and, as defined above, the communication time of device $k$ is $T_k^{\mathrm{tran}}$. Accordingly, the transmission capability can be expressed as 
\begin{equation}
    R_k=T_k^{\mathrm{tran}} B \log_2 \left(1 + \frac{p_k^\mathrm{c} |h_k^\mathrm{c}|^2}{\sigma_\mathrm{c}^2} \right),
\end{equation}
where $p_k^\mathrm{c}$, $h_k^\mathrm{c}$, $\sigma_\mathrm{c}^2$ represent 
the transmission power, channel gain, and channel noise power, respectively.

\subsection{Edge Computing Model}
The edge server receives the recovered feature $\bm{\tilde{w}}_k$ after the communication phase.
Classifiers, such as SVM and MLP, are employed to predict the class labels based on the concatenated feature vector $\bm{\tilde{w}} \triangleq \begin{bmatrix} \bm{\tilde{w}}_1^\top, \dots, \bm{\tilde{w}}_K^\top \end{bmatrix}^\top\!\! \in \mathbb{R}^D$ . Since the classification performance depends on the separability of feature vectors across different classes, we adopt the coding rate reduction function as an alternative metric to evaluate the sensing performance. The monotonic relationship between the sensing performance and the coding rate reduction will be demonstrated in Section~\ref{sec:converge}.

However, the coding rate reduction function introduced in Section~\ref{sys:feature} relies on the statistical properties of multiple samples and is therefore not directly applicable to single-sample edge inference scenarios.
In the following, we address this issue. According to~\cite{yu2020learning}, the concatenated feature vector $\bm{w}$ can be modeled as a real Gaussian mixture (GM) distribution, whose probability density function (PDF) can be denoted as 
\begin{equation}
p(\bm{w}) = \sum_{l \in \mathcal{L}} p_l \, \mathcal{N}^{\mathrm{v}}(\bm{w}; \bm{\mu}_l, \bm{\Sigma}_l, \bm{\Gamma}_l), 
\end{equation}
where $p_l$, $\bm{\mu}_l$, $\bm{\Sigma}_l$, and $\bm{\Gamma}_l$ denote the prior probability, mean vector, covariance matrix, and relation matrix of $\bm{w}$ in class $l$, respectively. Note that these parameters can be obtained via estimation 
over the training dataset.
Similarly, the recovered feature vector $\tilde{\bm{w}}$ at the BS also follows a GM distribution, and its PDF can be given by
\begin{equation}
p(\tilde{\bm{w}}) = \sum_{l \in \mathcal{L}} p_l \, \mathcal{N}^{\mathrm{v}}(\tilde{\bm{w}}; \bm{\mu}_l, \bm{\Sigma}_l+\bm{N}, \bm{\Gamma}_l+\bm{N}),
\end{equation}
where $\bm{N} \!\!=\!\! \mathrm{diag}(\bm{N}^{\mathrm{blk}}_1, \dots, \bm{N}_K^{\mathrm{blk}}) \!\!\in\!\! \mathbb{C}^{D \times D}$, and
$\bm{N}_k^{\mathrm{blk}} = \mathrm{diag} \left( \delta_{k,1}^2/Q_{k,1}, \dots, \delta_{k,d_k}^2/Q_{k,d_k} \right)$ 
represents the quantization distortion matrix for device $k$.  
Note that the coding rate function in equation~\eqref{equ:mcr2} is inherently determined by the sample covariance of the extracted features. Specifically, $\dfrac{1}{S}\bm{W} \bm{W}^\top$ in~\eqref{eq:R_W} and $\dfrac{1}{\mathrm{tr}(\bm{\Pi}_l)} \bm{W} \bm{\Pi}_l \bm{W}^\top$ in~\eqref{eq:Rc_W} represent the covariance matrices of the overall features (denoted by $\bm{\Sigma}$) and the features within class $l$ (i.e., $\bm{\Sigma}_l$), respectively.
Thus, by substituting $p_l$, $\bm{\Sigma}_l$, and 
$\bm{\Sigma}$ into equation~(\ref{equ:mcr2}), the coding rate reduction at the BS can be reformulated as
\begin{align}\label{equ:R_opti}
\Delta R(\bm{N}_k^{\mathrm{blk}}) 
 =& \log \det \left( \bm{I} + \alpha (\bm{\Sigma} + \bm{N}) \right) \nonumber \\
& - \sum_{l \in \mathcal{L}} p_l \log \det \left( \bm{I} + \alpha (\bm{\Sigma}_l\! +\! \bm{N}) \right), 
\end{align}
where $\alpha = \dfrac{D}{\epsilon^2}$. {It is worth noting that $\bm{\Sigma}$ and $\bm{\Sigma}_l$ are estimated from the concatenated multi-modal feature vectors obtained from a training dataset, and thus inherently encode cross-modal correlations through their off-diagonal blocks, which in turn affect the objective in~(\ref{equ:R_opti}). 
In a special case, when two modalities are highly correlated, they tend to contribute largely overlapping information, as reflected in the cross-modal covariance structure of $\bm{\Sigma}$ and $\bm{\Sigma}_l$.
As a result, their contributions to the MCR$^2$-based objective become highly similar. Consequently, under limited resources and delay requirements, optimizing the objective in~(\ref{equ:R_opti}) naturally avoids assigning excessive resources to redundant modal information. 
Overall, equation~(\ref{equ:R_opti}) characterizes the contribution of each feature to sensing performance based on the covariance matrices, and can therefore be adopted as the optimization objective in this paper.
}

\section{Problem Formulation and Transformation}\label{sec:reformulation}
In this section, we first formulate a sensing accuracy maximization problem and then transform it into an equivalent but more tractable form.
\subsection{Problem Formulation}
Our objective is to maximize the sensing accuracy,  which is equivalent to maximizing the coding rate reduction.
Mathematically, the optimization objective is expressed as\footnote{In this work, the sensing power $p_k^\mathrm{s}$ is treated as a pre-configured parameter, since we focus on system-level ISCC design after reliable sensing signals are already obtained. Optimizing $p_k^\mathrm{s}$ would require modeling its impact on feature extraction and the resulting MCR$^2$, which is left for future work.}
\begin{equation}
    \max\limits_{\left\{ Q_{k,i}, p_k^\mathrm{c}, T_k^{\mathrm{tran}}  \right\}} \Delta R(\bm{N}_k^{\mathrm{blk}}).
\end{equation}
Considering that the sensing task is delay-sensitive, the IoT devices are energy-constrained, and reliable data transmission must be ensured, we formulate the following three constraints.

\textit{1) Successful Transmission Constraint:} 
During the communication phase, to ensure that the quantized feature can be successfully transmitted to the edge server, the communication overhead for device $k$ must not exceed the transmission capability $R_k$, i.e., the following constraint should be satisfied
\begin{equation} \label{st:trans}
   \sum_{i=1}^{d_k} \log_2 \left(1 + \frac{Q_{k,i}}{\delta_k^2} \right) 
\leq T_k^{\mathrm{tran}} B \log_2 \left(1 + \frac{p_k^\mathrm{c} |h_k^\mathrm{c}|^2}{\sigma_\mathrm{c}^2} \right).
\end{equation}

\textit{2) Delay Constraint:}
The total delay of the sensing task should be limited. Let $T$ denote the upper limit, and, as analyzed in Section~\ref{ssec:iscc_model}, the delay constraint can be expressed as
\begin{equation}
    \max_k\{T_k^{\mathrm{sen}}+ T_k^{\mathrm{com}}\}+ \sum_{k=1}^{K} T_k^{\mathrm{tran}} \leq T.
\end{equation}

\textit{3) Energy Constraint:}
Due to the limited energy resources of IoT devices, the energy consumption of each device should be constrained as
\begin{equation}\label{st:E}   
p_k^\mathrm{s}T_k^{\mathrm{sen}}+E_k^{\mathrm{com}}+p_k^\mathrm{c}T_k^{\mathrm{tran}} \leq E_k,
\end{equation}
where $p_k^\mathrm{s}$ denotes the sensing power, $E_k^{\mathrm{com}}$ represents the energy consumed for local computation, and $E_k$ is the energy threshold for device $k$.

{From the above constraints, we can observe that the quantization gain, transmit power,
and allocated communication time are inherently coupled and jointly affect the sensing accuracy. Therefore, a joint optimization of these variables is necessary to improve the overall system performance.}
Specifically, the sensing accuracy maximization problem is formulated as
\begin{subequations} \label{pb_o}
\begin{eqnarray}
	& \max\limits_{\left\{ Q_{k,i}, p_k^\mathrm{c}, T_k^{\mathrm{tran}}  \right\}} &\Delta R(\bm{N}_k^{\mathrm{blk}}),\\
	&\text{s.t.} & \text{(\ref{st:trans})--(\ref{st:E})},\nonumber\\
    & & Q_{k,i}, p_k^\mathrm{c}, T_k^{\mathrm{tran}} \in \mathbb{R}^+.
\end{eqnarray}
\end{subequations}
Here, we do not explicitly consider the upper bounds on the quantization gain and communication transmit power, since the former is typically large and rarely reached in the considered resource-constrained scenario while the latter is already considered in the total energy budget. The upper bounds can be added as linear constraints without changing the problem structure, and the proposed algorithm in Section~\ref{sec:BCD} can be readily extended.

\begin{figure}[t]
	\setlength{\abovecaptionskip}{8pt} 
	\centering	
    \includegraphics[width=0.98\linewidth]{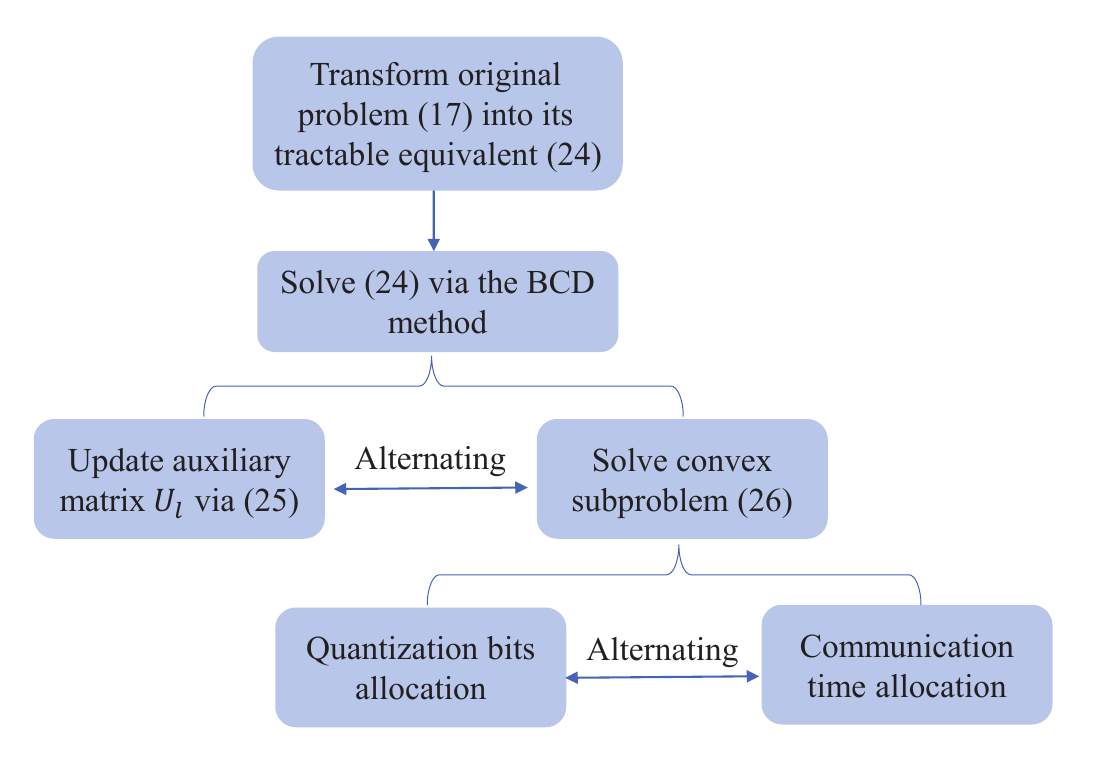}
	\vspace{-2ex}
	\caption{Solution procedure of the proposed ISCC framework.}
	\label{fig:alogtithm}
    \vspace{-2ex}
\end{figure}

Due to the non-convexity of both the objective function and constraint~(\ref{st:trans}), problem~(\ref{pb_o}) is a non-convex problem. To address this, we begin by simplifying the original problem, and the overall solution workflow is summarized in Fig.~\ref{fig:alogtithm}.

\subsection{Problem Transformation}
To simplify problem~(\ref{pb_o}), we first focus on the non-convex constraint and introduce two auxiliary variables as follows:
\begin{equation} \label{eq:vari}
    S_{k,i}=\frac{\delta_{k}^2}{Q_{k,i}}, \quad E_k^{\mathrm{tran}}=p_k^\mathrm{c}T_k^{\mathrm{tran}},
\end{equation}
where $S_{k,i}$ denotes the normalized quantization distortion, and $E_k^{\mathrm{tran}}$ represents the communication energy consumption for device $k$.
Then,
$\sum_{i=1}^{d_k}\log_2\left(1+\dfrac{Q_{k,i}}{\delta_k^2}\right)$ in constraint~\eqref{st:trans} can be equivalently rewritten as $\mathrm{Tr}\left(\log_2(\bm{I}+(\bm{N}_k^{\mathrm{blk}})^{-1})\right)$, where $\bm{N}_k^{\mathrm{blk}}=\mathrm{diag}\{S_{k,1},\ldots,S_{k,d_k}\}$, and the trace operator accounts for the summation of the element-wise logarithms over all feature dimensions. In the following, we use $\bm{N}_k^{\mathrm{blk}}$ to equivalently represent $Q_{k,i}$, providing a unified and convenient form for the subsequent optimization.
Moreover, $\log_2 \left(1 + \dfrac{p_k^\mathrm{c} |h_k^\mathrm{c}|^2}{\sigma_\mathrm{c}^2} \right)$ can be rewritten as $\log_2 \left(1 + \dfrac{E_k^{\mathrm{tran}} |h_k^\mathrm{c}|^2}{ T_k^{\mathrm{tran}} \sigma_\mathrm{c}^2} \right)$.
Then we focus on constraint~(\ref{st:E}), which can be transformed into
 \begin{equation} \label{st:E_max}
     E_k^{\mathrm{tran}} \leq E_k-p_k^\mathrm{s}T_k^{\mathrm{sen}}-E_k^{\mathrm{com}}.
 \end{equation}
To maximize the transmission capability , $E_k^{\mathrm{tran}}$ should
be taken as the maximum value that satisfies constraint~(\ref{st:E_max}), i.e., 
\begin{equation} \label{st:E_opti}
    E_k^{\mathrm{tran},\star}=E_k-p_k^\mathrm{s}T_k^{\mathrm{sen}}-E_k^{\mathrm{com}}.
\end{equation}
Based on above substitution, problem~(\ref{pb_o}) is converted into 
\begin{subequations} \label{pb_o1}
\begin{eqnarray}
&\!\!\!\max\limits_{\left\{ \bm{N}_k^{\mathrm{blk}},  T_k^{\mathrm{tran}}  \right\}} &\!\!\!\!\!\!\!  \log \det (\bm{I} + \alpha (\bm{\Sigma} + \bm{N})) \label{obj}\nonumber\\
   & & \!\!\!\!\!\!-\sum_{l \in \mathcal{L}} p_l \log \det \left( \bm{I} + \alpha (\bm{\Sigma}_l\! +\! \bm{N}) \right), \\
	&\text{s.t.} & \!\!\!\!\!\!\!\!\!\!\mathrm{Tr} \left( \log_2 \left( \bm{I} \!\!+\!\! (\bm{N}_k^{\mathrm{blk}})^{-1} \right) \right)\nonumber\\
& &\!\!\!\!\!\!\!\!\!\! \leq  T_k^{\mathrm{tran}} B \log_2\!\! \left(1 \!\!+\!\! \frac{E_k^{\mathrm{tran},\star} |h_k^\mathrm{c}|^2}{ T_k^{\mathrm{tran}} \sigma_\mathrm{c}^2} \right), \label{stt1}\\ 
    &   & \!\!\!\!\!\!\!\!\!\! {\max_k   \{T_k^{\mathrm{sen}}\! +\! T_k^{\mathrm{com}}\!\} +\! \sum_{k=1}^K T_k^{\mathrm{tran}}\! \leq \!T},\label{stt2}\\
    & & \!\!\!\!\!\!\!\!\!\!S_{k,i},  T_k^{\mathrm{tran}} \in \mathbb{R}^+. \label{stt5}
\end{eqnarray}
\end{subequations}
In problem~(\ref{pb_o1}), all constraints are convex, but the objective function is still non-concave. The proof for convexity can be found in Appendix~\ref{proof:conv_con}.
For the objective function, we have the following lemma for transforming it.

\begin{lem}\label{lem:U}
  For any positive definite matrix \( \bm{F} \in \mathbb{C}^{r \times r} \), the following relation holds: 
\begin{equation} \label{eq:tranform}
     - \log \det (\bm{F}) = \max_{\bm{U} \succ 0} \{\log \det (\bm{U}) - \operatorname{tr} (\bm{U} \bm{F})\} + r,
\end{equation}
where $\bm{U}$ is a positive definite matrix acting as an auxiliary variable. The right-hand side of (\ref{eq:tranform}) is maximized at
\begin{equation} \label{eq:U_max}
    \bm{U}^{\star} = \bm{F}^{-1}.
\end{equation}
    \begin{IEEEproof}
    Since~(\ref{eq:tranform}) is an unconstrained convex problem, we can find the optimal solution by taking the first-order derivative in terms of $\bm{U}$ to zero. Substituting~(\ref{eq:U_max}) back into~(\ref{eq:tranform}), the equality in~(\ref{eq:tranform}) is verified. The detailed derivation is omitted due to the page limit.
\end{IEEEproof}
\end{lem}
Based on Lemma~\ref{lem:U},
problem~(\ref{pb_o1}) can be further converted into the following problem
\begin{eqnarray}\label{pb_o2}
	\!\!\!\!\!\!\!\!\!\!\!\!&\max\limits_{\small\left\{ \bm{N}_k^{\mathrm{blk}}, \bm{U}_l,   T_k^{\mathrm{tran}}  \right\}} & \!\!\!\!\!\!\!\!\log \det (\bm{I}\!\! + \alpha (\bm{\Sigma}\!\! +\!\! \bm{N}))\nonumber\\
    \!\!\!\!\!\!\!\!\!\!\!\!&&\!\!\!\!\!\!\!\! +  \sum_{l \in \mathcal{L}} p_l  \log \det (\bm{U}_l)   \nonumber \\
    &&\!\!\!\!\!\!\!\!\!\!- \!\sum_{l \in \mathcal{L}} p_l\mathrm{tr} (\bm{U}_l ( \bm{I} + \alpha (\bm{\Sigma}_l + \bm{N}))), \\
	\!\!\!\!\!\!\!\!\!\!\!\!&\text{s.t.} & \!\!\!\!\!\!\!\!\!\!\!\!\text{(\ref{stt1})--(\ref{stt5})}.\nonumber
\end{eqnarray}
Problem~(\ref{pb_o2}) is equivalent to~(\ref{pb_o1}) in that the optimal solution $\bm{N}^{\star}$ is identical for both problems.
In the following, we adopt the BCD~\cite{BCD} method to address this problem.

\section{Sensing Accuracy Maximization}\label{sec:BCD}

In this section, we first decompose problem~\eqref{pb_o2} into two subproblems and develop a BCD-based algorithm.

\subsection{Problem Decomposition}
After problem reformulation in Section~\ref{sec:reformulation}, we propose a BCD-based algorithm to solve~(\ref{pb_o2}), which updates one block of variables at a time while keeping the other blocks fixed. 
Specifically, there are two blocks, i.e., $\bm{U}_l$ and $\{\bm{N}_k^{\mathrm{blk}}, T_k^{\mathrm{tran}}\}$. 
Under fixed $\bm{N}_k^{\mathrm{blk}}$ and $T_k^{\mathrm{tran}}$, problem~(\ref{pb_o2}) is an unconstrained convex optimization problem with $\bm{U}_l$. From Lemma~\ref{lem:U}, we can obtain the optimal solution
\begin{equation}\label{eq:U_opti}
    \bm{U}^*_l=( \bm{I} + \alpha (\bm{\Sigma}_l\!\! + \!\!\bm{N}))^{-1}.
\end{equation}
For fixed $\bm{U}_l$, problem~(\ref{pb_o2}) can be reformulated as
\begin{eqnarray}\label{pb_omain}
	&\max\limits_{\left\{ \bm{N}_k^{\mathrm{blk}},   T_k^{\mathrm{tran}}  \right\}} &\!\!\!\!\log \det (\bm{I}\!\! + \!\! \alpha (\bm{\Sigma}\!\! +\!\! \bm{N}))\!\! \nonumber\\
    &&-\sum_{l \in \mathcal{L}} \alpha p_l  \mathrm{tr} (\bm{U}_l \bm{N}),\\
	&\text{s.t.} & \text{(\ref{stt1})--(\ref{stt5})}.\nonumber
\end{eqnarray}
Problem~(\ref{pb_omain}) is a convex problem, as the objective function consists of a concave term and a linear term, and all constraints are convex. However, jointly optimizing all variables still leads to prohibitive computational complexity of $\mathcal{O}((\sum_k d_k+K)^3)$, which grows cubically with the number of devices $K$. This high complexity is primarily attributed to the high-dimensional matrix optimization and the tight coupling between $\bm{N}_k^{\mathrm{blk}}$ and $T_k^{\mathrm{tran}}$. To address this, we adopt an alternating optimization (AO) strategy to reduce the overall computational burden.
 
\subsection{An Alternating Algorithm for Solving~(\ref{pb_omain})}
{In this subsection, we develop an AO strategy by decomposing~(\ref{pb_omain}) into 
the quantization bit allocation\footnote{In practice, the quantization distortion is determined by the number of bits allocated to each feature, and these two terms actually reflect the same underlying optimization variables. Therefore, we use ``quantization bit allocation'' to identically represent the optimization of $\bm{N}_k^{\mathrm{blk}}$.} and communication time allocation subproblems. Specifically, we first initialize $T_k^{\mathrm{tran}}$ to  obtain the corresponding optimal $\bm{N}_k^{\mathrm{blk}}$. We next compute the minimum total transmission delay $\sum_k T_k^{\mathrm{tran}}$ to verify the feasibility of the current quantization levels, which is then used to guide the communication time reallocation. This process is repeated until convergence. 
Finally, a convergence analysis of the proposed AO strategy is provided.}
  
\subsubsection{Quantization Bit Allocation}
In this case, the communication time $T_k^{\mathrm{tran}}$ is fixed.
Problem~(\ref{pb_omain}) can be converted into the following problem
\begin{eqnarray}\label{pb_oq}
	&\max\limits_{\left\{ \bm{N}_k^{\mathrm{blk}}  \right\}} &\log \det (\bm{I} +  \alpha (\bm{\Sigma} + \bm{N})) \nonumber\\
    &&-\sum_{l \in \mathcal{L}} \alpha p_l  \mathrm{tr} (\bm{U}_l \bm{N}),\\
	&\text{s.t.} & \text{(\ref{stt1}) and (\ref{stt5})}. \nonumber
\end{eqnarray}
 Since the computational complexity of solving problem~(\ref{pb_oq}) scales with the dimension of the optimization matrix $\bm{N}$, and constraint~(\ref{stt1}) is separable across devices, we optimize the block variables ${\bm{N}_k^{\mathrm{blk}}}$ sequentially device by device. For each device, the resulting subproblem reduces to a low-dimensional convex program, which can be efficiently solved using standard convex optimization techniques, such as projected gradient descent or interior-point methods.

\subsubsection{Communication Time Allocation}
In this case, the quantization distortion matrix $\bm{N}$ is fixed and $T_k^{\mathrm{tran}}$ does not appear at the objective function in problem~(\ref{pb_omain}). Thus, the communication time allocation reduces to finding a feasible set of $T_k^{\mathrm{tran}}$ that satisfies~(\ref{stt1}) and (\ref{stt2}) simultaneously. On this basis, we further select the solution that enables more efficient optimization in the next iteration.

{
To proceed, we first examine whether a feasible solution exists under the current quantization. Accordingly, we compute the minimum total transmission delay of all devices while satisfying constraint~(\ref{stt1}), thereby formulating the following problem.
}
\begin{subequations} \label{pb_o5}
\begin{eqnarray}
	&\min\limits_{\left\{  T_k^{\mathrm{tran}}  \right\}} & \sum_{k=1}^K T_k^{\mathrm{tran}}, \\
	&\text{s.t.} & \!\!\!\!\mathrm{Tr} \left( \log_2 \left( \bm{I} \!\!+\!\! (\bm{N}_k^{\mathrm{blk}})^{-1} \right) \right)\nonumber\\
& &\!\!\!\!\leq  T_k^{\mathrm{tran}} B \log_2\!\! \left(1 \!\!+\!\! \frac{E_k^{\mathrm{tran},\star} |h_k^\mathrm{c}|^2}{ T_k^{\mathrm{tran}} \sigma_\mathrm{c}^2} \right), \label{stt11}\\ 
    & & \!\!\!\!T_k^{\mathrm{tran}} \in \mathbb{R}^+. 
\end{eqnarray}
\end{subequations}
It can be easily verified that the problem~(\ref{pb_o5}) is convex, 
as the objective function is linear and the constraint is convex. 
Moreover, for each device $k$, the left-hand side of~(\ref{stt11}), denoted as $C_k$, is a constant, whereas the right-hand side is a continuous and strictly increasing function of $T_k^{\mathrm{tran}}$. Hence, the optimal communication time $ T_k^{\mathrm{tran},\star}$ can be obtained when constraint~(\ref{stt11}) holds with equality. This value can be efficiently obtained using a bisection search, as summarized in Algorithm~\ref{alg:time}.
Since $T_k^{\mathrm{sen}}$ and $T_k^{\mathrm{com}}$ are fixed constants for each device, the term 
$\max_k\{T_k^{\mathrm{sen}}+ T_k^{\mathrm{com}}\}$ can be regarded as a constant, denoted as $T^{\mathrm{c}}$.
Then, to ensure feasibility,
the upper and lower bounds of $ T_k^{\mathrm{tran},\star}$ are given by
\begin{align}
\left \{
    \begin{array}{ll}
        T_k^{\mathrm{tran},\star}\geq T_k^{\min}=0,\\
		  T_k^{\mathrm{tran},\star}\leq T_k^{\max}=T-T^{\mathrm{c}} .
\end{array} 
\right.
\end{align}

{\begin{algorithm}[t]
		\caption{Communication Time Allocation for Solving~({\ref{pb_o5}}).}
		\label{alg:time}
		\DontPrintSemicolon  
            \textbf{Input:} $\bm{N}_k^{\mathrm{blk}}$, $E_k^{\mathrm{tran},\star}$ \;
        Set the maximal error tolerance $\epsilon>0$;\;
        Set 
        $T_k^{\mathrm{l}}=T_k^{\min}, T_k^{\mathrm{u}}=T_k^{\max}$;\;
        
        \Repeat{$|C_k -R_k|<\epsilon$;}
        {
        Let $T_k^{\mathrm{tran}, \star}=(T_k^{\mathrm{l}}+T_k^{\mathrm{u}})/2$;\;
        Calculate $R_k=T_k^{\mathrm{tran},\star} B \log_2\!\! \left(1 \!\!+\!\! \frac{E_k^{\mathrm{tran},\star} |h_k^\mathrm{c}|^2}{ T_k^{\mathrm{tran},\star} \sigma_\mathrm{c}^2} \right)$; \;
        \eIf{$C_k > R_k$}{
        $T_k^{\mathrm{l}} = T_k^{\mathrm{tran}, \star}$;\;  
        }{
            $T_k^{\mathrm{u}} = T_k^{\mathrm{tran}, \star}$;\;
        }
       
        }
      \textbf{Output:} $T_k^{\mathrm{tran}, \star}$.
\end{algorithm} 
}

From Algorithm~\ref{alg:time}, we obtain the minimal {total transmission delay} required under the current quantization, i.e., $\sum_{k=1}^K T_k^{\mathrm{tran},\star}$.
We then define the remaining delay budget as
\begin{equation}
        \Delta T= T-(T^{\mathrm{c}}+\sum_{k=1}^K T_k^{\mathrm{tran},\star}).
\end{equation}
Then, we have the following three cases.
\begin{itemize}
    \item When $\Delta T>0$, the current quantization configuration is feasible, and the additional communication time can be reallocated to IoT devices to further improve the quantization gain.

    \item When $\Delta T<0$, the delay constraint is violated, indicating that the quantization bits must be reduced in the next iteration\footnote{A feasible initialization could prevent the algorithm from entering this case.}.

    \item When $\Delta T=0$, the current time allocation is optimal.
\end{itemize}

When $\Delta T$ is nonzero, we further design a redistribution scheme to allocate the remaining or exceeding time among devices, as
\begin{equation} \label{eq:updateT}
    T_{k}^{\mathrm{tran}} = T_{k}^{\mathrm{tran}, \star} + \frac{ \eta_k }{ \sum_{k=1}^{K} \eta_k } \times \Delta T, \quad \forall k,
\end{equation}
where  the communication time reallocation weight $\eta_k$ is set to be the communication overhead of device $k$, i.e., 
\begin{equation}\label{eq:weight}
\eta_k = \mathrm{Tr} \left( \log_2 \left( \mathbf{I} + (\bm{N}_k^{\mathrm{blk}})^{-1} \right) \right) , \quad \forall k.
\end{equation}    

The update rule in~(\ref{eq:updateT}) is well justified.  When surplus communication time is available, the reallocation rule allocates more time to devices that require higher communication throughput, thereby improving resource utilization and potentially enhancing the objective function.
Conversely, when no feasible solution exists, the update rule guides the algorithm toward restoring feasibility in the next iterations.

\subsubsection{Convergence of the Alternating Algorithm}
The complete procedure for solving problem~(\ref{pb_omain}) is summarized in Algorithm~\ref{alg:AO}.
Before analyzing the convergence, we first give the following theorem.
\begin{thm}\label{thm:time}
     With increased $T_{k}^{\mathrm{tran}}$, the quantization distortion can be reduced, which in turn enhances the coding rate reduction and thereby improves the sensing accuracy.
\begin{IEEEproof}
	Please refer to Appendix~\ref{proof:theo1}.
\end{IEEEproof}
\end{thm}
Theorem~\ref{thm:time} ensures that the alternating updates of the quantization bits and communication time allocation lead to a monotonic improvement in the coding rate reduction objective.
Moreover, since problem~(\ref{pb_omain}) is convex, the alternating optimization procedure converges to the global optimum with a linear rate, as indicated in~\cite{2019federated,wen2023task}.

{\begin{algorithm}[t]
		\caption{Alternating Algorithm for Solving~({\ref{pb_omain}}).}
		\label{alg:AO}
		\DontPrintSemicolon  
		\textbf{Initialize:} Communication time $T_k^{\mathrm{tran}}$;\;
        \Repeat{\textnormal{Convergence} }
        {
        \For{$k=1,2,\cdots, K$}{
         \%Quantization bits allocation;\;
         Solve problem~(\ref{pb_oq}) using CVX and attain${\bm{N}_k^{\mathrm{blk}}}$;\;
         \%Time allocation strategy;\;
         Solve problem~(\ref{pb_o5}) using Algorithm~\ref{alg:time};\;
        Update $T_k^{\mathrm{tran}}$ using~(\ref{eq:updateT});

         }        
         
		}       
      \textbf{Output:} $\bm{N}_k^{\mathrm{blk}}$, $T_k^{\mathrm{tran}}$.
\end{algorithm} 
}

\subsection{Overall Algorithm}\label{sec:overall_alo}
Based on the above analysis, the overall algorithm for solving problem~(\ref{pb_o1}) is summarized in Algorithm~\ref{alg:overall}. The proposed method follows a BCD structure consisting of two blocks.
In the first block, the auxiliary matrices $\bm{U}_l$ are updated in closed form. In the second block, the communication resources are allocated using the AO strategy.
After the BCD iterations converge, 
the obtained solution is substituted back through the transformations in~(\ref{eq:vari}), from which the optimal solution to problem~(\ref{pb_o}) is recovered.
In each iteration, updating the $\bm{U}_l$ requires a computational complexity of $\mathcal{O}(D^3)$. Then, we derive the quantization bits and time allocation strategy by alternately solving subproblems~(\ref{pb_oq}) and~(\ref{pb_o5}).
The complexity of solving~(\ref{pb_oq}) can be calculated as $\mathcal{O}(\sum_{k=1}^Kd_k^3)$, and solving~(\ref{pb_o5}) requires $\mathcal{O}(K\mathrm{log} (T_k^{\max}))$. Let $I_{0}$ and $I_{1}$ denote the maximum iteration numbers of Algorithm~\ref{alg:AO} and Algorithm~\ref{alg:overall}, respectively. Then the total computational complexity of Algorithm~\ref{alg:overall} is $\mathcal{O}(I_1(D^3+I_0\sum_{k=1}^Kd_k^3+I_0K\mathrm{log}(T_k^{\max})))$.
Since each block admits an optimal solution and the overall procedure follows a standard BCD framework, the convergence of the proposed algorithm is guaranteed~\cite{BCD}.

\begin{figure}[t]
	\setlength{\abovecaptionskip}{8pt} 
		\centering		\includegraphics[width=0.98\linewidth]{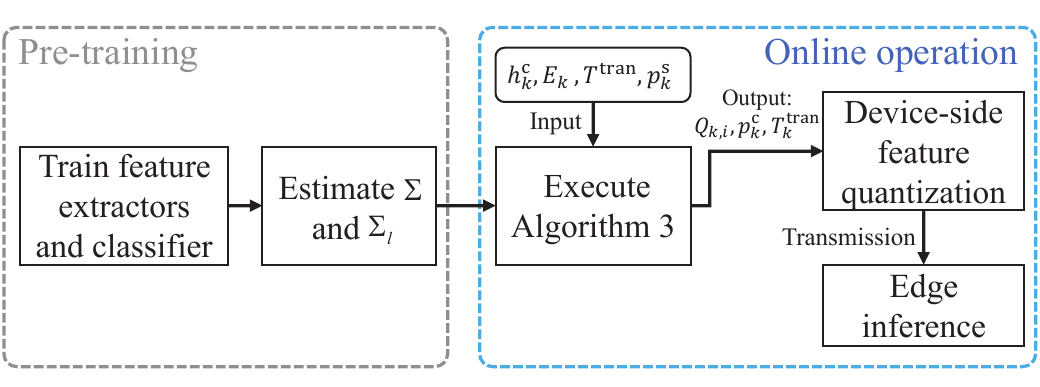}
		\vspace{-2ex}
		\caption{The workflow of the proposed system.}
		\label{fig:workflow}
        \vspace{-2ex}
\end{figure}

{The detailed workflow of the considered system in practical deployment is illustrated in Fig.~\ref{fig:workflow}.
Before the online execution stage, the edge server first trains the feature extractors and classifiers using a pre-collected dataset. The trained feature extractors are then applied to this dataset to characterize the feature distribution, from which $\bm{\Sigma}$ and $\bm{\Sigma}_l$ are obtained.  
During online operation, the devices report their channel conditions, energy constraints, delay requirements, and other parameters to the BS. Based on these parameters, the BS runs Algorithm~\ref{alg:overall} to determine the communication time allocation, quantization distortion matrix, and power configuration, which are then delivered to the devices for actual feature compression and transmission, followed by edge inference at the BS.
}

{\begin{algorithm}[t]
		\caption{BCD-based Algorithm for Solving~({\ref{pb_o}}).}
		\label{alg:overall}
		\DontPrintSemicolon  
		\textbf{Initialize:} Quantization distortion matrix $\bm{N}$;\;
        \Repeat{\textnormal{Convergence} }
        {
        Update $\bm{U}_l$ according to~(\ref{eq:U_opti});\;   
        Solve~(\ref{pb_omain}) under given $\bm{U}_l$
        using Algorithm~\ref{alg:AO} and get $\bm{N}_k^{\mathrm{blk}}$,  $T_k^{\mathrm{tran}}$;\;
        
		}
        Calculate $Q_{k,i}$ and $p_k^\mathrm{c}$ according to~(\ref{st:E_opti}) and~(\ref{eq:vari});\;
      \textbf{Output:} $Q_{k,i}$, $p_k^\mathrm{c}$, $T_k^{\mathrm{tran}}$.
\end{algorithm} 
}

\section{Test Results}\label{sec:test_all}
In this section, we conduct experiments to verify the effectiveness of the proposed algorithm.

\subsection{Test Setup}\label{sec:test}

The test settings described below are applied unless otherwise specified. 
In the test, we consider the human activity recognition task and adopt a multi-model dataset, namely XRF55~\cite{wang2024xrf55}, for multi-modal edge inference.
The dataset consists of three modalities, i.e., Wi-Fi, RFID, and mmWave radar, and it contains 55 human activity types. Among them, we select eight representative types: waving, clapping hands, falling on the floor, jumping, sitting down, standing up, turning, and walking. Each action type contains 600 instances, yielding a total of 4,800 samples, of which 3,360 samples are used for training and the remaining 1,440 samples are reserved for testing.
For feature extraction, each device is equipped with a ResNet18~\cite{resnet} backbone. These networks are jointly optimized offline under the loss function defined in~\eqref{sec:obj_feature} over a Linux server equipped with two NVIDIA GeForce GTX 4090 GPUs, and subsequently deployed to the devices. Moreover,
the statistical parameters of the extracted features, i.e., $\bm{\Sigma},\bm{\mu}_l,\bm{\Sigma}_l$, and $\bm{\Gamma}_l$, are estimated by averaging the feature samples over the training dataset.
For edge inference, two machine learning classifiers are employed, i.e., SVM and MLP neural network. Both models are trained on 3,360 feature samples without quantization distortion, while testing is conducted on 1,440 distorted feature samples that are distributed uniformly across the eight classes to evaluate sensing accuracy.


After training the ResNet18 and classifiers and obtaining the feature statistics, we then simulate the communication process.
{We consider an ISCC system with $K=3$ ISAC devices and the radius of the BS's coverage is 500 meters~\cite{wang2024joint}.  
Following the 3rd Generation Partnership Project (3GPP) channel modeling guidelines~\cite{zhu20213gpp}, we incorporate both large-scale and small-scale fading in the wireless channel model.
Specifically, the large-scale path loss is modeled as $128.1+37.6\log_{10}d$, where $d$ denotes the distance in kilometers, and the large-scale shadowing is modeled as a Gaussian distribution with zero mean and standard deviation 8~dB~\cite{zhao2024iov,liu2024optimal}. The small-scale fading is assumed to follow a Rayleigh distribution with unit variance. 
Unless otherwise specified, the remaining simulation parameters are summarized in Table~\ref{tab:sim_params}, following commonly adopted settings in the ISCC literature~\cite{wang2023device,he2023integrated,liu2025latency}}. 
It is worth clarifying that the sensing time $T_k^{\mathrm{sen}}=1~\!\mathrm{s}$ in Table~\ref{tab:sim_params} represents the observation window required to collect sufficient sensing samples for recognizing an activity, rather than continuous occupation of the wireless channel. During this period, the sensing device performs intermittent signal transmissions/receptions at a certain sampling rate. In addition, the local computation time $T_k^{\mathrm{com}}=0.01~\!\mathrm{s}$ is adopted as a conservative upper bound for local feature extraction, ensuring that even the slowest device can complete local processing within this duration. Nevertheless, 0.01~\!s is only an example used in our simulation while other values can be also used for tests.
Based on the parameter settings above and the trained models, the optimized results obtained from Algorithm~\ref{alg:overall} are used to perform the overall simulation following the procedures described in Section~\ref{ssec:iscc_model}.
{Moreover, although MCR$^2$ serves as an alternative optimization objective, we report the experimental results in terms of sensing accuracy, as it is the most direct and practically relevant measure for the human activity recognition task.}

\begin{table}[t]
\centering
\caption{Simulation Parameters}
\label{tab:sim_params}
\resizebox{\columnwidth}{!}{%
\vspace{-1ex}
\begin{tabular}{c c}
\toprule
\textbf{Parameter} & \textbf{Value} \\
\hline \hline
Sensing time, $T_k^{\mathrm{sen}}$ & 1~\!s \\ \hline
Computation time, $T_k^{\mathrm{com}}$ & 0.01~\!s  \\ \hline
Permitted {total transmission delay}, $T^{\mathrm{tran}}$ & 0.09~\!s  \\ \hline
Energy threshold, $E_k$ & 0.16~\!Joule \\ \hline
Computation energy, $E_k^{\mathrm{com}}$ & 0.03~\!Joule  \\ \hline
Sening power, $p_k^{\mathrm{s}}$ & 0.04,0.02,0.06~\!W  \\ \hline
Quantization distortion variance, $\delta_k^2$ & 1  \\ \hline
Communication noise power, $\sigma_\mathrm{c}^2$ & -174~\!dBm/Hz  \\ \hline
Feature dimension per device, $d_k$ & 20 \\ \hline
Bandwidth for communication, $B$ & 170~\!KHz  \\ \hline
Coding rate distortion level, $\epsilon$ & 0.1 \\ 
\bottomrule
\end{tabular}%
}
\end{table}

\subsection{Metric Validation and Algorithm Investigation}\label{sec:converge}

{This subsection 
conducts micro-benchmark experiments to evaluate the relations between the MCR$^2$ metric and the sensing accuracy, and to analyze the convergence and computation delay behavior of the proposed BCD-based algorithm.}

\begin{figure}[t]
	\setlength{\abovecaptionskip}{8pt} 	
	\centering
	\includegraphics[width=0.7\linewidth]{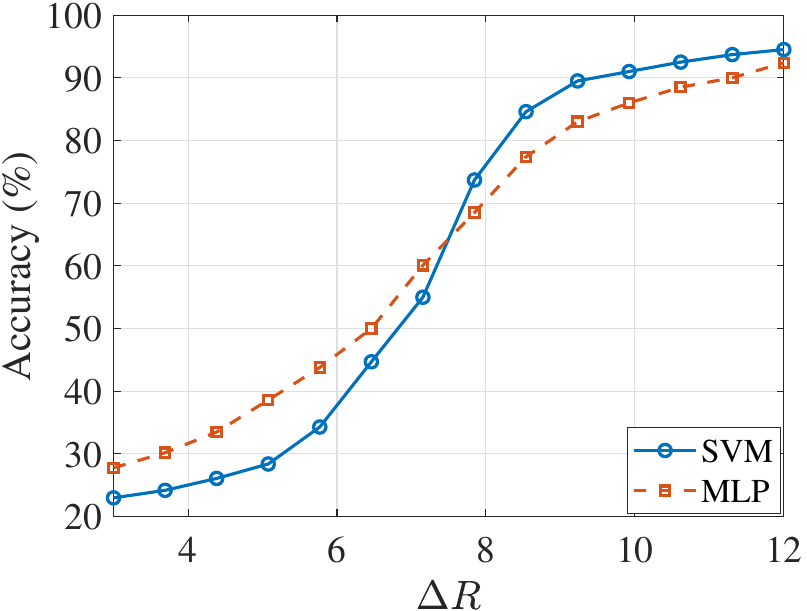}
	\vspace{-1ex}
    \caption{Sensing accuracy with different $\Delta R$.}
	\label{fig:acc_mcr2}
    \vspace{-2ex}
\end{figure}

\textbf{Effectiveness of the MCR$^2$ metric}. Fig.~\ref{fig:acc_mcr2} illustrates the relation between the sensing accuracy and the $\mathrm{MCR}^2$ metric for both the SVM and MLP models.
{Each point corresponds to one quantization distortion matrix $\bm{N}$ generated by sweeping the distortion level, from which the corresponding coding rate reduction value is computed. The sensing accuracy is then evaluated on the $\bm{N}$-distorted test samples using the pre-trained SVM or MLP models.}
It can be observed that the sensing accuracy monotonically increases as $\Delta R$ grows for both models, indicating that the $\mathrm{MCR}^2$ objective provides a suitable alternative measure to evaluate the sensing accuracy. 
Moreover, when $\Delta R$ is small, i.e., the feature quantization distortion is large, the MLP model outperforms the SVM model, due to its stronger robustness to noise in high-distortion conditions. However, when $\Delta R$ is large, i.e., the distortion is small, the SVM model achieves a higher accuracy. This result comes from the fact that a well-trained MLP model tends to be more sensitive to slight distortions compared to the margin-based SVM model.

\begin{figure}[t]
    \vspace{-1ex}
	\centering
	\includegraphics[width=0.98\linewidth, trim = 0 0 0 15, clip]{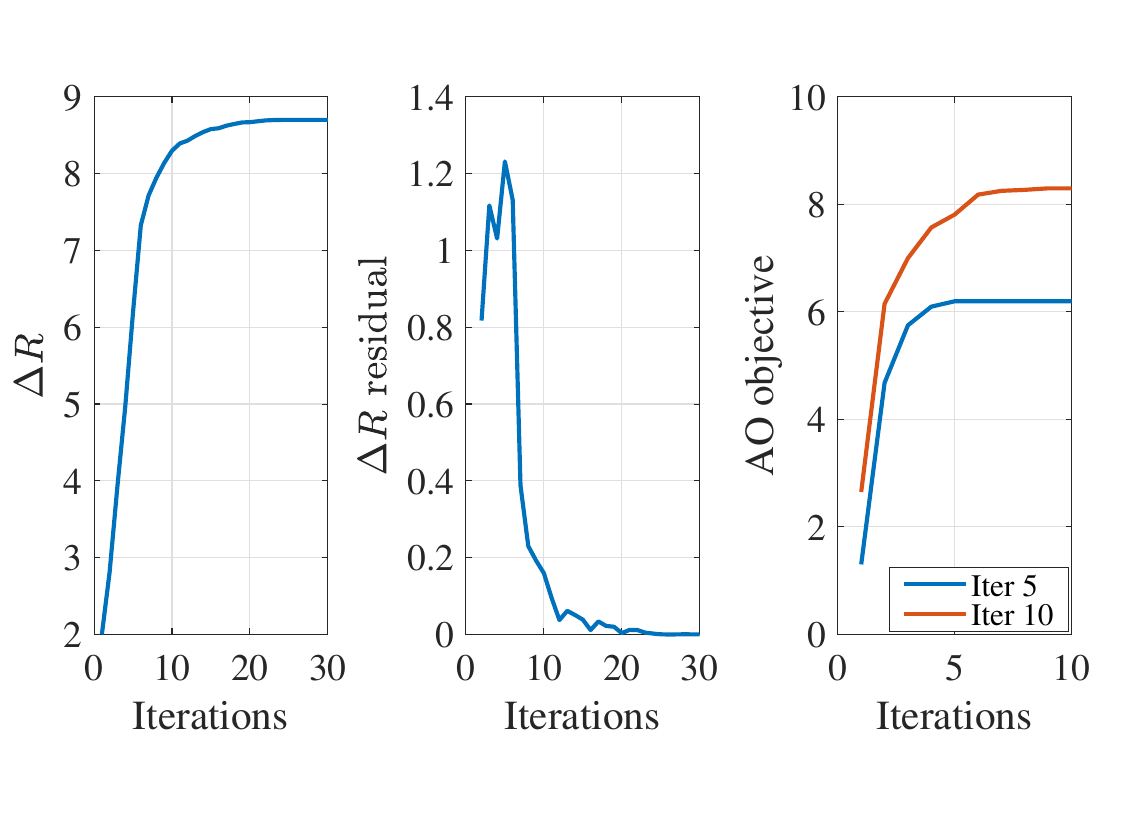}
	\vspace{-4ex}
    \caption{Convergence behavior of the proposed algorithms. The left two subfigures illustrate the convergence of the overall BCD-base algorithm, while the right subfigure shows the convergence of the inner AO procedure, where ``Iter 5'' and ``Iter 10'' correspond to the 5th and 10th  BCD iterations, respectively. }
	\label{fig:converge}
    \vspace{-2ex}
\end{figure}

\textbf{Convergence behavior}. Fig.~\ref{fig:converge} depicts the convergence behavior of the proposed BCD-based algorithm under parameters described in Section~\ref{sec:test}. As shown in the left two subfigures, the overall algorithm converges within approximately 20 BCD iterations, and the objective residual decreases rapidly to zero, demonstrating a fast convergence rate. Moreover, the right subfigure compares the AO convergence trajectories at the 5th and 10th BCD  iterations.
In both cases, the AO objective stabilizes within about 10 iterations, indicating the consistent and efficient convergence across different stages of the BCD procedure.


{\textbf{Computation delay analysis.}
Due to the high-level software overhead in our simulations, the measured running time of the proposed algorithm when implemented in MATLAB is not representative of an optimized edge server implementation.
To assess practical deployability, we provide an engineering-oriented theoretical delay estimate. 
Under the parameter settings in Section~\ref{sec:test}, the total computational cost is approximately 74~\!MFLOPs based on the computational complexity analyzed in Section~\ref{sec:overall_alo}.
Assuming a conservative effective throughput of 20~\!GFLOPs/s (e.g., on an Intel Xeon Bronze–class CPU platform), the estimated computation delay is 3.70~\!ms, which is practically feasible for real-time tasks. Table~\ref{tab:latency_dk} further reports how the delay grows as the per-device feature dimension increases, thereby illustrating the scalability of the proposed method. Moreover, the resource allocation optimization is updated at a slower timescale than human activity recognition inference, which further facilitates real-time operation.}

\begin{table}[t]
\centering
\caption{Estimated computation delay versus per-device feature dimension $d_k$.}
\label{tab:latency_dk}
\begin{tabular}{ c  c c c c c }
\toprule
$d_k$ & 20 & 25 & 30 & 35 & 40 \\
\hline
Delay~\!(ms) & 3.70 & 6.81 & 11.31 & 17.46 & 25.53 \\
\bottomrule
\end{tabular}
\end{table}

\subsection{Performance Evaluation} \label{sec:perform}
{This subsection evaluates the proposed algorithm through comparisons with an existing semantic communication method and several system-level baselines, as well as an experiment examining the effect of cross-modal correlation on the resource allocation results.}

{\subsubsection{Comparisons with an Existing Semantic Communication Method} Our objective may appear to be related to that of some semantic communication works, but we should note that the underlying design considerations are fundamentally different. To further clarify our contribution and validate the effectiveness of the proposal, we compare it with a representative semantic communication scheme.
Specifically, we adopt the deep joint source-channel coding (JSCC) method in~\cite{bourtsoulatze2019deep} as the baseline, and consider the additive white Gaussian noise (AWGN) channel. For a fair comparison, both schemes are configured to satisfy the same end-to-end delay requirement, including the sensing, computation, and communication stages. To ensure this latency consistency, we adjust the model size of the baseline while keeping its original framework unchanged. In this baseline, semantic features are
first extracted by the encoder and then quantized before transmission. Moreover, since this baseline does not incorporate a dedicated resource-allocation mechanism, the available transmission bits under a given resource budget are uniformly allocated across devices and feature components in our implementation.
The quantized features are then transmitted to the edge side over the AWGN channel, and the recovered features are subsequently fed into the decoder for task inference.

We report the comparative results under two representative system settings, where the permitted transmission delay is set to 0.03~\!s and 0.09~\!s, respectively, as shown in Fig.~\ref{fig:JSCC}. One can clearly see that the proposed scheme achieves higher sensing accuracy under both settings.
A possible explanation is as follows.
The proposed scheme focuses on system-level ISCC design and resource allocation, where the MCR$^2$ metric inherently captures cross-modal correlations and guides the allocation of transmission resources toward modalities/features that provide more complementary information for the sensing task. Therefore, the proposed scheme tends to be more advantageous in resource-constrained multi-modal sensing scenarios.
In contrast, the main purpose of the JSCC baseline is to learn an end-to-end joint source-channel coding architecture for efficient feature compression and transmission over an AWGN channel.
 Therefore, the two approaches do not share exactly the same design objective. Nevertheless, the above results still demonstrate the advantage of the proposed scheme in the considered multi-modal ISCC system. 

\begin{figure}[t]
	\setlength{\abovecaptionskip}{8pt} 
		\centering		\includegraphics[width=0.7\linewidth]{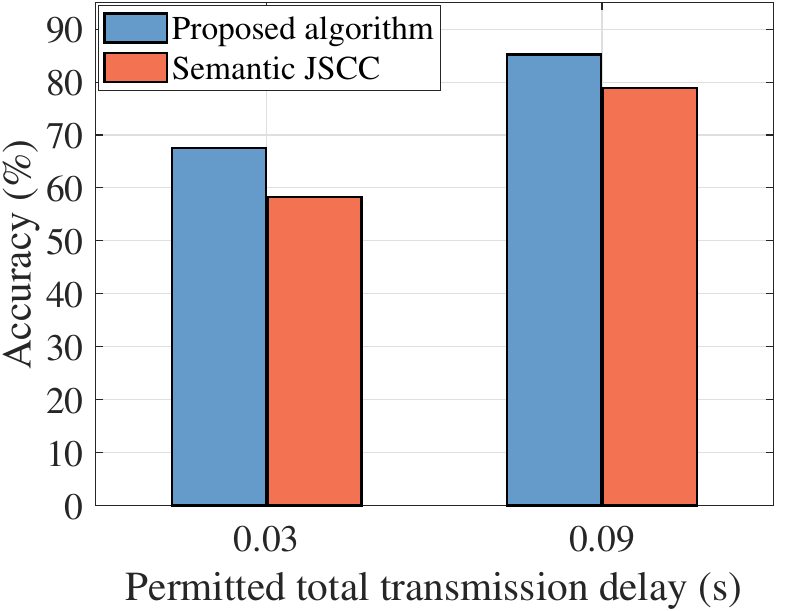}
		\vspace{-1ex}
		\caption{Performance comparison with the semantic JSCC baseline. The proposed algorithm adopts the SVM classifier.}
		\label{fig:JSCC}
        \vspace{-1ex}
\end{figure}
}
{
\subsubsection{Comparisons with System-Level Baselines}
To further demonstrate the effectiveness of the proposed algorithm and the benefit of multi-modal sensing, we consider the following baselines.}

\begin{itemize}
    \item Device-level quantization scheme: 
Each device applies the same quantization bit to all its features, i.e., all elements of $\bm{N}_k^{\mathrm{blk}}$ are constrained to be equal. This is equivalent to adding a constraint of $\bm{N}_k^{\mathrm{blk}}=n_k\bm{I}$ to problem~(\ref{pb_o1}), and the new problem can be solved following a similar manner as Algorithm~\ref{alg:overall}.
\item Average time allocation scheme: The communication time is first equally allocated to each device and the remaining parameters are obtained using
Algorithm~\ref{alg:overall}.
\item Single modality scheme: Only the WiFi modality is used for sensing, as it is the most widely adopted modality in wireless sensing~\cite{li2024reshaping,he2023sencom,luo2024vision}. {In this case, the feature dimension is also set to 24, which is empirically determined as the optimal value through training experiments. Moreover, all available communication resources are exclusively allocated to this modality.}
\end{itemize}

\begin{figure}[t]   
    \centering
    \subfigure[SVM]{
       \centering
        \includegraphics[width=0.7\linewidth]{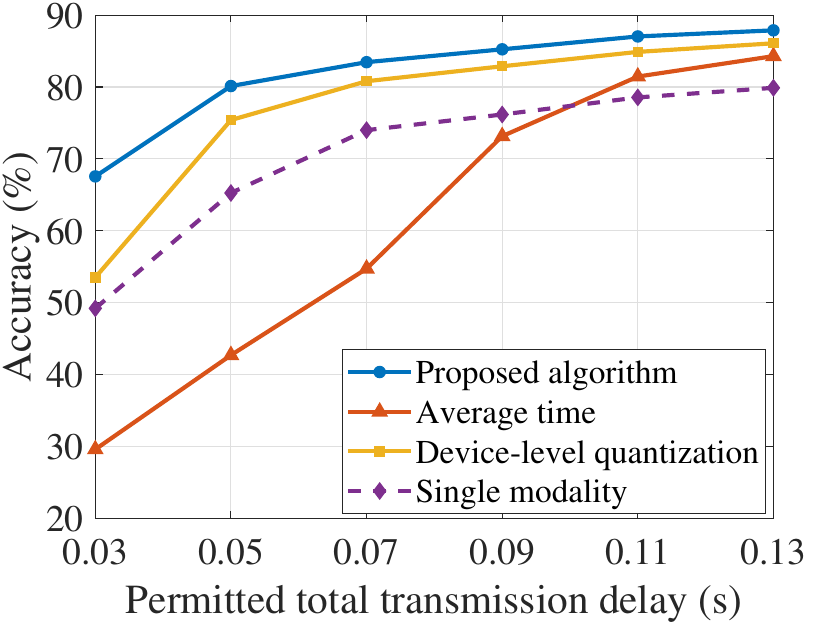}\label{fig:acc_T_SVM}
    }
    \subfigure[MLP]{
    \centering
    \includegraphics[width=0.7\linewidth]{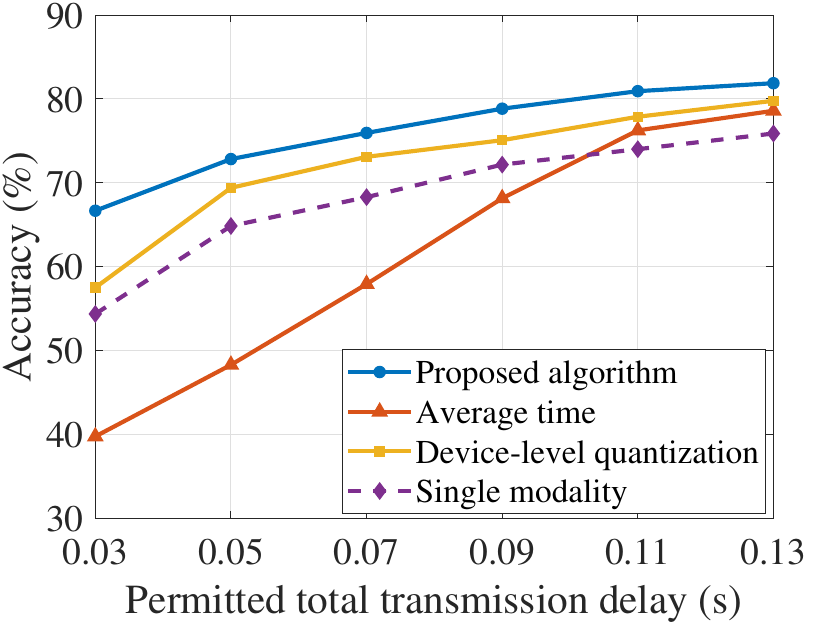}\label{fig:acc_T_MLP}
    }  
    \caption{Sensing accuracy vs. permitted {total transmission delay}.} 
    \label{fig:acc_T}
    \vspace{-2ex}
\end{figure}

\textbf{Overall performance}. 
Figs.~\ref{fig:acc_T_SVM} and~\ref{fig:acc_T_MLP} illustrate the impact of the permitted {total transmission delay} on the sensing accuracy for four schemes, using the SVM and MLP models, respectively. It can be seen that, for all schemes, the sensing accuracy improves as the delay budget increases. 
This can be attributed to that the transmission capability  is positively correlated with the communication time.
As the permitted delay grows, each device can be allocated more communication time, which enhances the transmission capability 
and enables a lower quantization distortion that
still satisfies the successful transmission constraint. Moreover, the proposal consistently outperforms both the average time allocation and the device-level quantization schemes, especially under tight delay budgets, demonstrating the superiority of the proposed resource allocation strategy. When the delay constraint becomes relatively loose, the single modality scheme yields the lowest accuracy due to its limited information content,  further validating the advantage of multi-modal sensing. 
Since the MLP and SVM models exhibit similar performance trends, subsequent results are presented using the SVM model as a representative example.

\begin{figure}[t]
	\setlength{\abovecaptionskip}{8pt} 	
	\centering
	\includegraphics[width=0.7\linewidth]{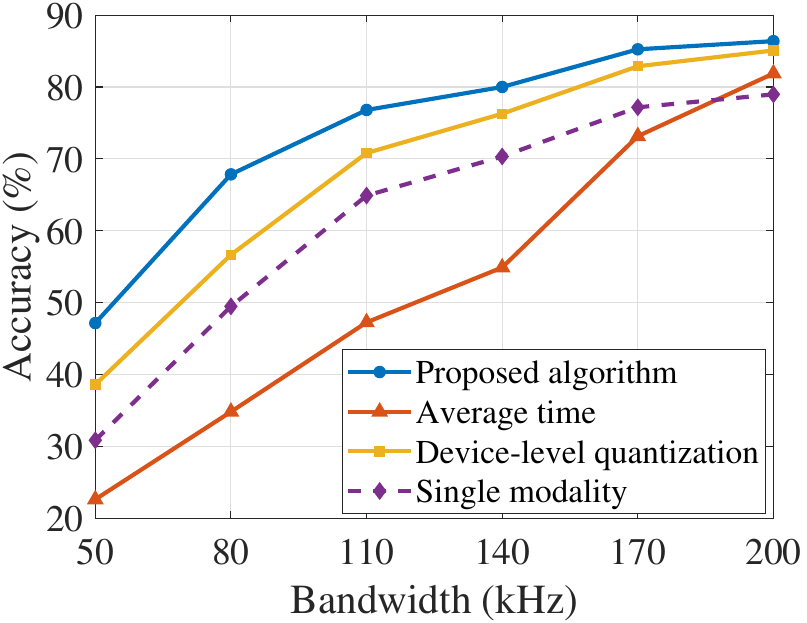}
	\vspace{-1ex}
	\caption{Sensing accuracy vs. communication bandwidth.}
	\label{fig:acc_B}
    \vspace{-1ex}
\end{figure}

\textbf{Effect of the communication bandwidth}. 
Fig.~\ref{fig:acc_B} depicts the impact of the communication bandwidth on the sensing accuracy for four schemes. Again, our proposed scheme demonstrates the best performance under varying bandwidth. Besides, when the bandwidth is small, the sensing accuracy of all schemes increases rapidly as the available bandwidth grows. This is because, in the low-bandwidth regime, the transmission capability  grows approximately linearly with the bandwidth, thereby effectively mitigating quantization distortion and improving sensing accuracy. As the bandwidth continues to increase, the marginal gain in sensing accuracy gradually diminishes. Once the bandwidth becomes sufficiently large, the quantization distortion of all schemes is greatly reduced, thereby narrowing the performance gap between the proposed scheme and the baselines.

\begin{figure}[t]
	\setlength{\abovecaptionskip}{8pt} 	
	\centering
	\includegraphics[width=0.7\linewidth]{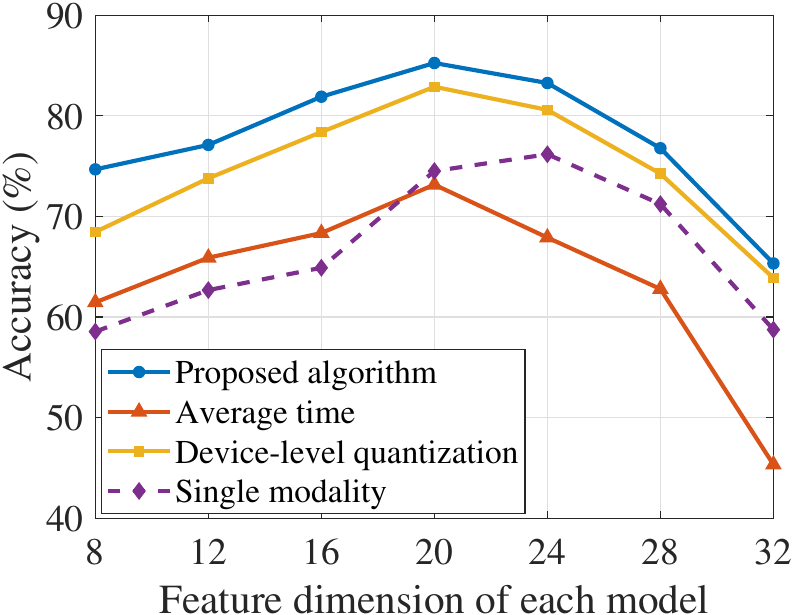}
	\vspace{-1ex}
	\caption{Sensing accuracy vs. feature dimension of each model.}
	\label{fig:acc_dim}
    \vspace{-1ex}
\end{figure}

\textbf{Effect of the feature dimension}. 
Fig.~\ref{fig:acc_dim} shows the effect of the feature dimension of each modality on the sensing accuracy for four schemes. For all schemes, the accuracy first increases and then decreases as the feature dimension grows.  The reason behind it can be explained as follows. When the feature dimension is small, increasing it enhances the representational capability of the extracted features, thereby improving the sensing accuracy. However, when the feature dimension becomes too large, two adverse effects emerge.
On the one hand, the extracted features may contain redundant or noise-dominated components that contribute little to the sensing performance. On the other hand, under limited communication resources, a higher feature dimension leads to fewer quantization bits per feature, which in turn amplifies the quantization distortion. These combined effects eventually cause a noticeable degradation in sensing accuracy.

\begin{figure}[t]
	\setlength{\abovecaptionskip}{8pt} 	
	\centering
	\includegraphics[width=0.7\linewidth]{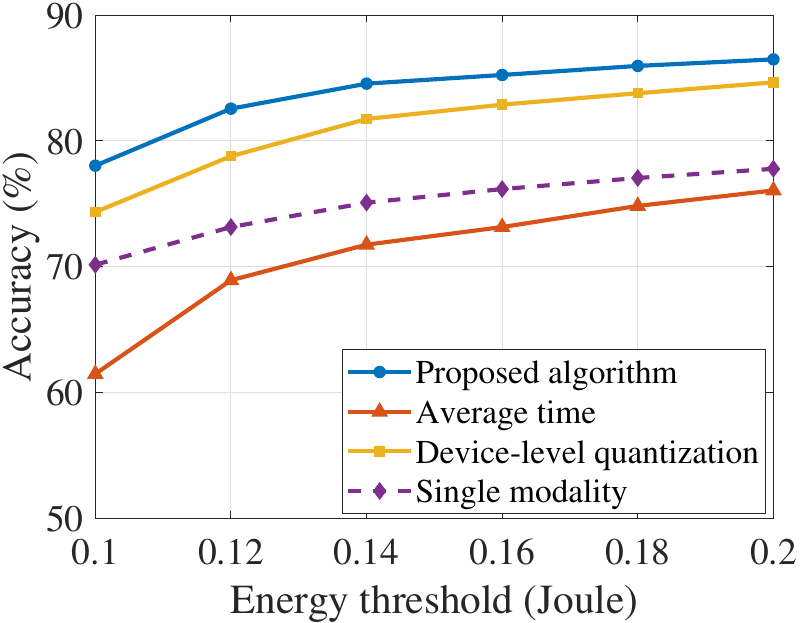}
	\vspace{-1ex}
	\caption{Sensing accuracy vs. energy threshold.}
	\label{fig:acc_E}
    \vspace{-1ex}
\end{figure}

\textbf{Effect of the energy threshold}. 
Fig.~\ref{fig:acc_E} demonstrates the impact of the energy threshold on the sensing accuracy for four schemes. As observed, the performance of all schemes improves with the energy threshold.
With a larger energy budget, each device can either extend its communication time or increase its transmit power, thereby enhancing the transmission capability  and mitigating the quantization distortion, which ultimately leads to improved sensing accuracy. 
Moreover, the device-level quantization scheme consistently achieves lower sensing accuracy than the proposed scheme, indicating that different features contribute unequally to the sensing performance and validating the effectiveness of feature-level quantization.

\begin{figure}[t]
	\setlength{\abovecaptionskip}{8pt} 	
	\centering
    \includegraphics[width=0.7\linewidth]{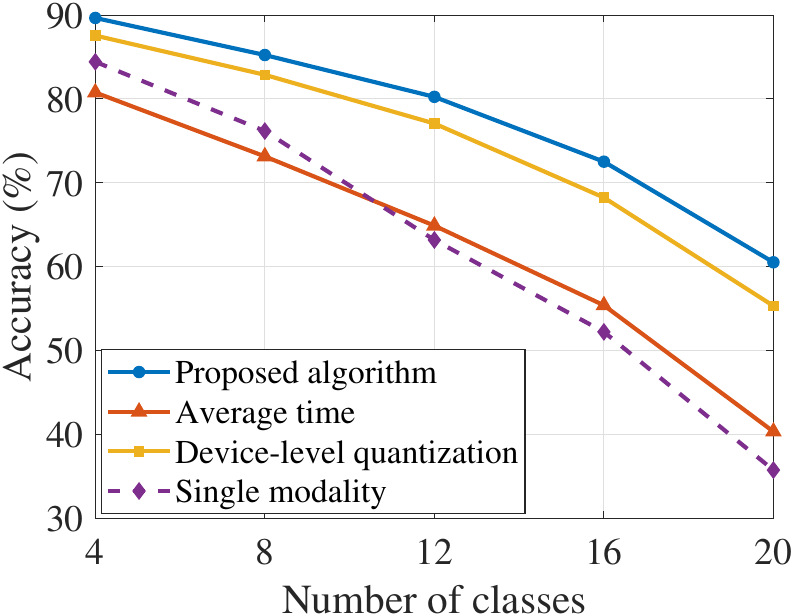}
	\vspace{-1ex}
	\caption{Sensing accuracy vs. number of classes.}
	\label{fig:acc_type}
    \vspace{-1ex}
\end{figure}

\textbf{Effect of the number of classes}. 
Fig.~\ref{fig:acc_type} shows the effect of the number of classes on the sensing accuracy for four schemes. We can observe that the accuracy of all schemes decreases with the number of classes. This trend is reasonable, since a larger number of classes tightens the inter-class margins and complicates the decision boundaries, while the fixed-dimensional quantized features offer limited separability. The proposed scheme consistently achieves the highest accuracy across the entire range, and its advantage becomes more pronounced as the task becomes more challenging (i.e., with more classes), highlighting the effectiveness of the proposed resource allocation strategy. Moreover, the performance of the single-modality scheme degrades the fastest. When the number of classes exceeds 12, its accuracy falls below that of the average time allocation scheme, further confirming the superiority of multi-modal sensing.

{\subsubsection{Effect of Cross-Modal Correlation on the Resource Allocation} We further conduct a supplementary experiment to demonstrate the effect of cross-modal correlation on the resource allocation results. 
To this end, we consider a setting where all three devices use the same modality, namely, the WiFi modality. In addition, two different channel configurations are considered. The corresponding results are reported in Table~\ref{tab:time_alloc}. Under channel setting 1, the channel gains are ranked as device~2 $>$ device~1 $>$ device~3, whereas under channel setting 2, the ranking becomes device~3 $>$ device~1 $>$ device~2. As can be seen from the table, when all devices use the same modality, the allocated resources are positively correlated with the channel quality, that is, a device with a better channel is allocated more resources. This is reasonable because the features provided by different devices are essentially substitutable in this case, and a better channel allows more information to be delivered, thereby improving the sensing performance more effectively. In contrast, under the original multi-modal setting with three different modalities, more resources are allocated to device~3, which corresponds to the mmWave modality. This is because the mmWave beam is narrower and more concentrated on the sensing target, and therefore provides more informative features. As a result, even when its channel condition is relatively weaker, the system still tends to allocate more resources to the mmWave modality. 

}

\begin{table}[h]
\centering
\caption{Communication-time allocation under two different settings.}
\label{tab:time_alloc}
\resizebox{\columnwidth}{!}{%
\begin{tabular}{c c c c c}
\toprule
\textbf{Channel} & \textbf{Modality} & \textbf{Device 1} & \textbf{Device 2} & \textbf{Device 3} \\
\hline \hline
\multicolumn{1}{c|}{\multirow{2}{*}{\makecell{Setting 1}}}
& Multi-modal & 0.021\!~s & 0.018\!~s & 0.051\!~s \\ \cline{2-5}
\multicolumn{1}{c|}{}
& All-WiFi & 0.028\!~s & 0.041\!~s & 0.021\!~s \\ \hline
\multicolumn{1}{c|}{\multirow{2}{*}{\makecell{Setting 2}}}
& Multi-modal & 0.018\!~s & 0.014\!~s & 0.058\!~s \\ \cline{2-5}
\multicolumn{1}{c|}{}
& All-WiFi & 0.029\!~s & 0.028\!~s & 0.033\!~s \\ \hline
\end{tabular}%
}
\end{table}

\section{Conclusion}\label{sec:conclu}
In this paper, we studied a task-oriented multi-modal ISCC framework that combines local feature extraction at IoT devices with joint multi-modal inference at the edge server. The MCR$^2$ criterion was employed to guide local feature extraction, thereby reducing task execution delay and avoiding raw-data transmission. Then, the influence of quantization distortion, transmit power, and communication capacity
on the system performance was quantitatively analyzed. To further improve the performance of the multi-modal ISCC system under limited resources, we formulated a sensing accuracy  maximization problem using the MCR$^2$ metric under delay and energy constraints. Subsequently, the resulting nonconvex problem was transformed into an equivalent tractable form, and a BCD-based algorithm was developed to obtain an efficient solution.
Finally, extensive experiments on publicly available datasets validated that the proposed framework and algorithm consistently achieve higher sensing accuracy than all baseline schemes, highlighting the benefits of multi-modal sensing and coordinated resource allocation.

\appendices

\section{Proof of Convexity for Constraints in~\eqref{pb_o1}}\label{proof:conv_con}
The first constraint in~(\ref{pb_o1})
is given by
\begin{equation}
    \mathrm{Tr} \left( \log_2 \left( \bm{I} \!\!+\!\! (\bm{N}_k^{\mathrm{blk}})^{-1} \right) \right) \leq  T_k^{\mathrm{tran}} B \log_2\!\! \left(1 \!\!+\!\! \frac{E_k^{\mathrm{tran}} |h_k^\mathrm{c}|^2}{ T_k^{\mathrm{tran}} \sigma_\mathrm{c}^2} \right).
\end{equation}
The left-hand side is a convex function, since its second derivative is positive. The right-hand side can be regarded as a linear transformation of $f(x,y)=x\log_2(1+y/x)$, which is concave because its Hessian matrix is negative semidefinite. Since concavity is preserved under linear transformations, the right-hand side of the first constraint is also concave. Therefore, the first constraint is a convex constraint.
The second and third constraints are linear, and are therefore convex as well. Thus, all constraints are convex, which ends the proof.

\section{Proof of Theorem~\ref{thm:time}}\label{proof:theo1}

In problem~(\ref{pb_omain}), constraint~(\ref{stt1}) is
\begin{equation}
    \mathrm{Tr} \left( \log_2 \left( \bm{I} \!\!+\!\! (\bm{N}_k^{\mathrm{blk}})^{-1} \right) \right)\leq  T_k^{\mathrm{tran}} B \log_2\!\! \left(1 \!\!+\!\! \frac{E_k^{\mathrm{tran}} |h_k^\mathrm{c}|^2}{ T_k^{\mathrm{tran}} \sigma_\mathrm{c}^2} \right).
\end{equation}
It can be observed that the right-hand side of constraint is monotonically increasing with respect to the communication time $T_k^{\mathrm{tran}}$. 
Therefore, by allocating a longer communication time, the transmission capability  on the right-hand side becomes larger. 
As a result, the constraint can still be satisfied with a smaller quantization distortion  $\bm{N}_k^{\mathrm{blk}}$, i.e., higher quantization gain can be achieved when $T_k^{\mathrm{tran}}$ increases.
Next, the objective function of problem~(\ref{pb_omain}) is 
\begin{equation}
    \log \det (\bm{I}\!\! + \!\! \alpha (\bm{\Sigma}\!\! +\!\! \bm{N}))
    -\sum_{l \in \mathcal{L}}\alpha p_l \mathrm{tr}(\bm{U}_l \bm{N}).
\end{equation}
When the quantization noise $\bm{N}$ decreases, the effective covariance matrix $\bm{\Sigma}+\bm{N}$ exhibits larger eigenvalues, which in turn increases the value of the log-determinant term $\log \det (\bm{I}+\alpha(\bm{\Sigma}+\bm{N}))$. In addition, the second term $\sum_{l \in \mathcal{L}}\alpha p_l \mathrm{tr}(\bm{U}_l \bm{N})$, depends linearly on $\bm{N}$ and decreases as $\bm{N}$ becomes smaller. Taken together, these two effects ensure that the overall objective function increases monotonically with reduced quantization distortion, which ends the proof.

\bibliographystyle{IEEEtran}
\bibliography{ref}

\begin{IEEEbiography}[{\includegraphics[width=1in,clip]{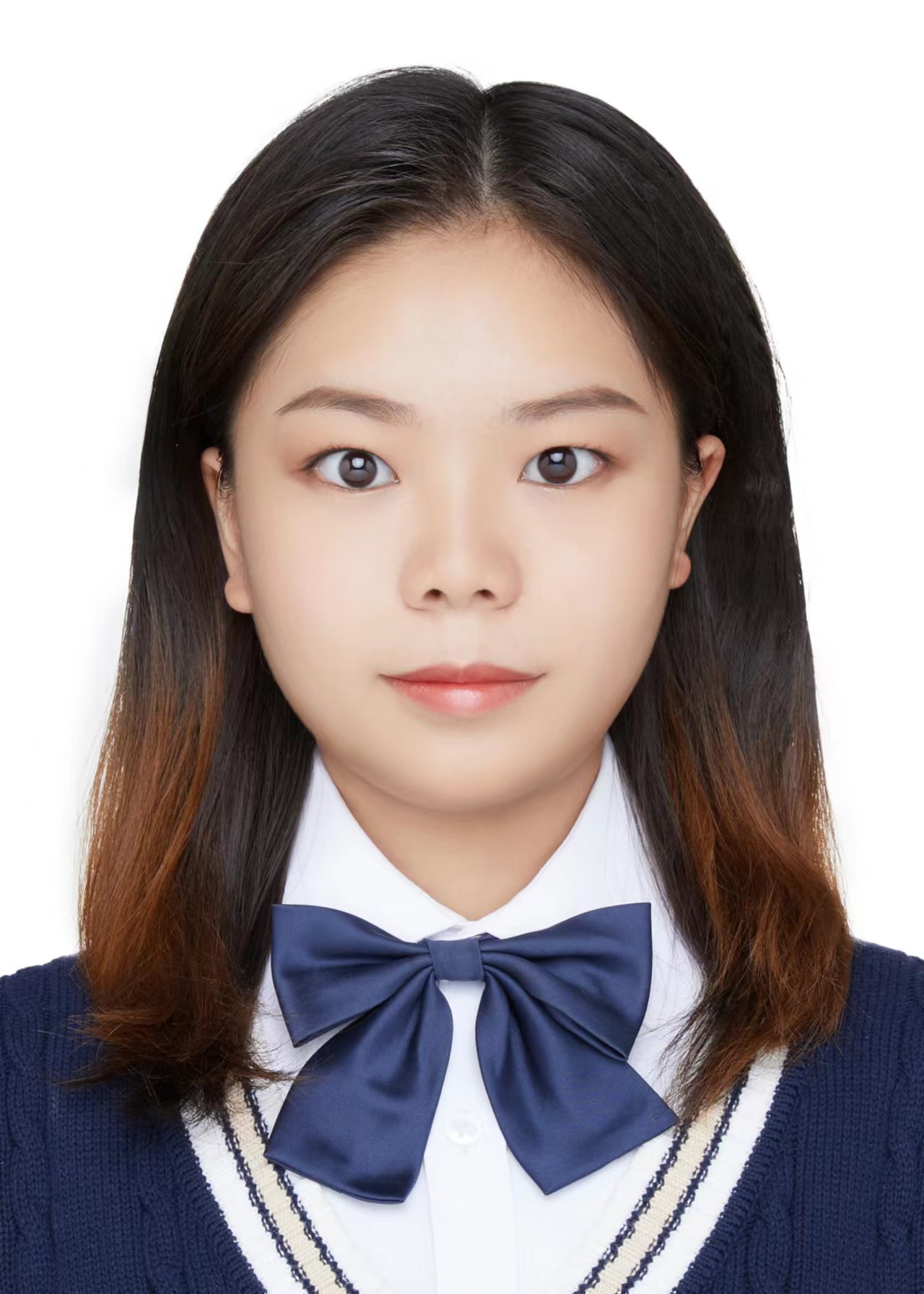}}]{Weiwei Chen} received the B.E. degree in Communication Engineering from Beijing University of Posts and Telecommunications, Beijing, China, in 2023. She is currently pursuing the M.E. degree in the College of Information Science and Electronic Engineering at Zhejiang University, Hangzhou, China. Her research interests mainly include integrated sensing and communication and edge inference.
\end{IEEEbiography}

\begin{IEEEbiography}	
	[{\includegraphics[width=1in,clip,keepaspectratio]{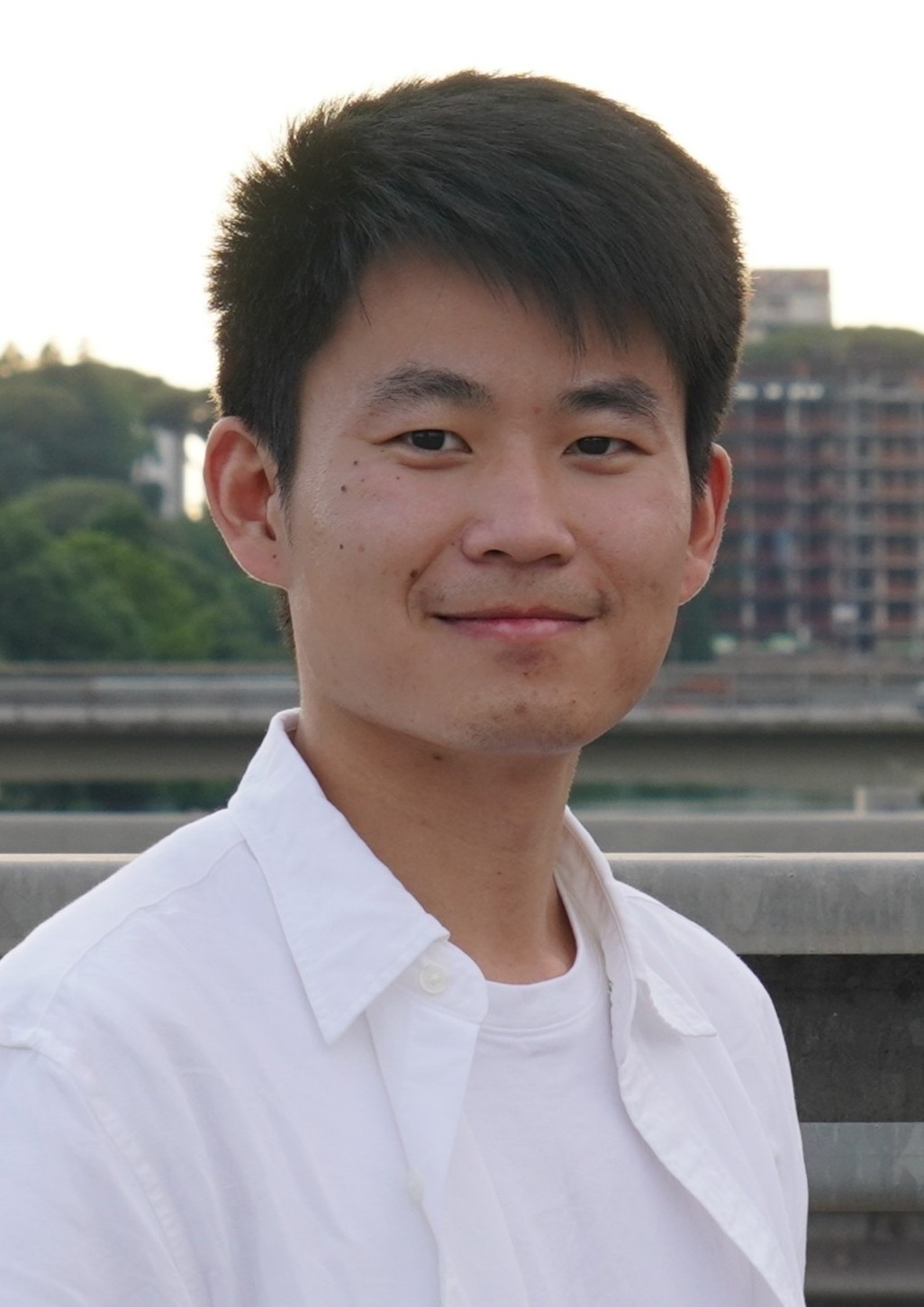}}] 
	{Yinghui He} (Member, IEEE) received the B.E. degree in information engineering and Ph.D. degree in information and communication engineering from Zhejiang University, Hangzhou, China, in 2018 and 2023, respectively. 
    He is currently a Research Fellow with the College of Computing and Data Science, Nanyang Technological University. 
    His research interests mainly include integrated sensing and communications (ISAC), mobile computing, and device-to-device communications.
\end{IEEEbiography}

\begin{IEEEbiography}	
	[{\includegraphics[width=1in,clip,keepaspectratio]{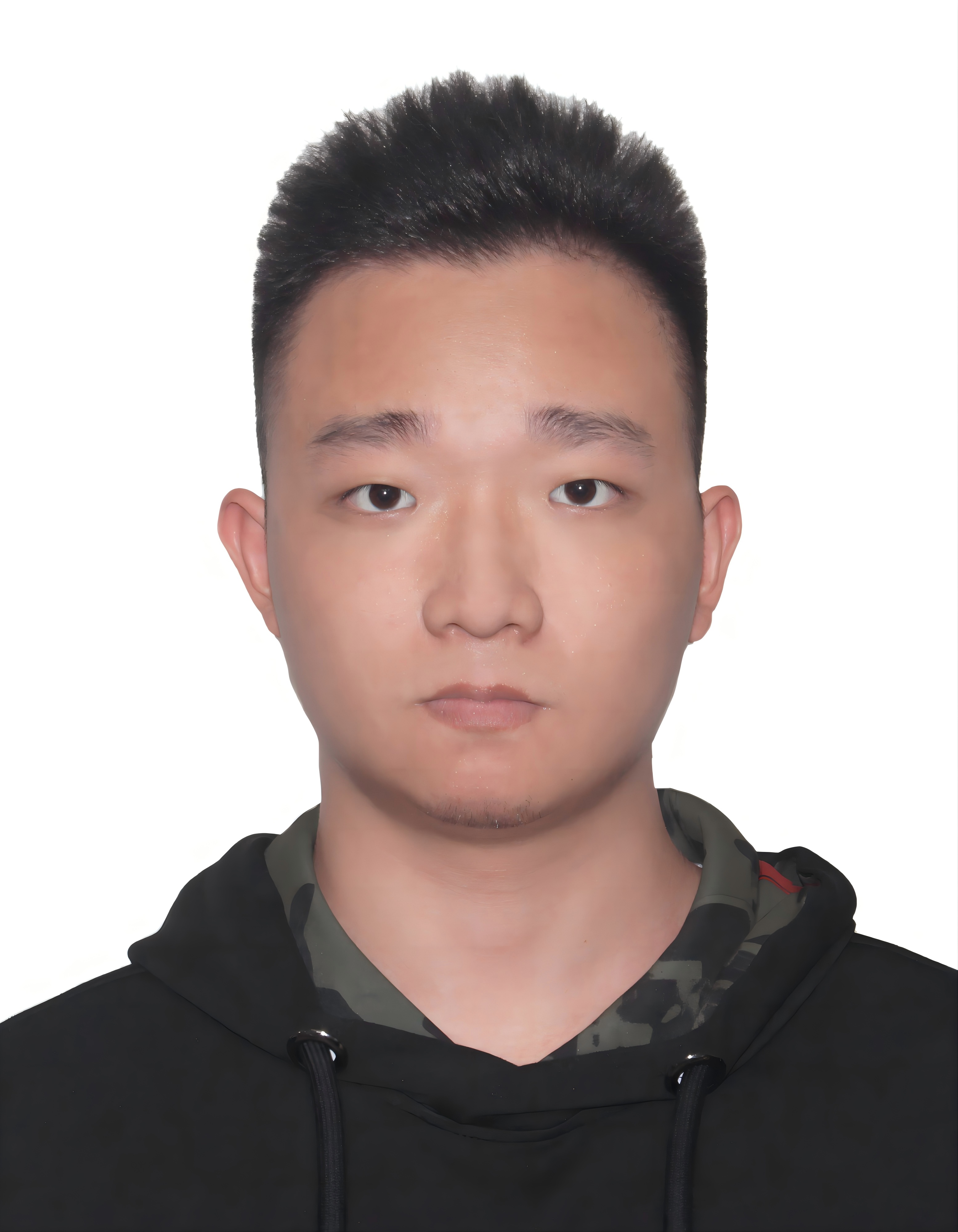}}] 
	{Zhong Ye} received the B.E. degree in communication engineering form Zhejiang University City College, Hangzhou, China, in 2019, and the M.Sc degree in information and communication engineering from Zhejiang Gongshang University, Hangzhou, China. He is currently pursuing the Ph.D. degree at Zhejiang University, and his research interests mainly include vehicular networking systems, multimodal sensing fusion and integrated sensing and communications.
\end{IEEEbiography}

\begin{IEEEbiography}	
	[{\includegraphics[width=1in,clip,keepaspectratio]{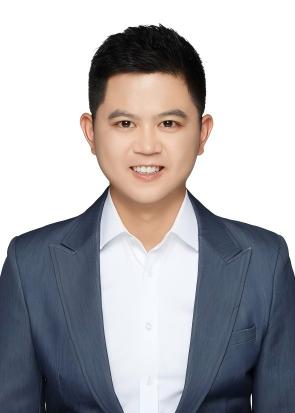}}] 
	{Dingzhu Wen} (Member, IEEE) is an Assistant Professor at the School of Information Science and Technology in ShanghaiTech University. He received his Bachelor's and Master's degrees from the Department (School) of Information Science and Electronic Engineering of Zhejiang University in 2014 and 2017, respectively. He received his Ph. D. degree from the Department of Electrical and Electronic Engineering of The University of Hong Kong in 2021. His research interests include brain-computer communication, edge AI, task-oriented communications, and integrated sensing-communication-computation. He has served as a co-organizer for workshops at flagship IEEE conferences including ICC, GlobeCom, WCNC, PIMRC, and VTC, and as a tutorial co-organizer at GlobeCom, WCNC, ICCC, and PIMRC. He has also chaired technical sessions at IEEE ICC, VTC, and WCSP. He was named as the Exemplary Reviewer for IEEE Transactions on Communications (2022), IEEE Transactions on Mobile Computing (2025), and IEEE Transactions on Network Science and Engineering (2025). He was awarded the IEEE GlobeCom 2023 Workshop Best Paper. He was selected as a World's Top 2\% Scientist (Stanford University \& Elsevier) and received the Excellent Mentor Award from ShanghaiTech University in 2023 and 2025.
\end{IEEEbiography}

\begin{IEEEbiography}	
	[{\includegraphics[width=1in,clip,keepaspectratio]{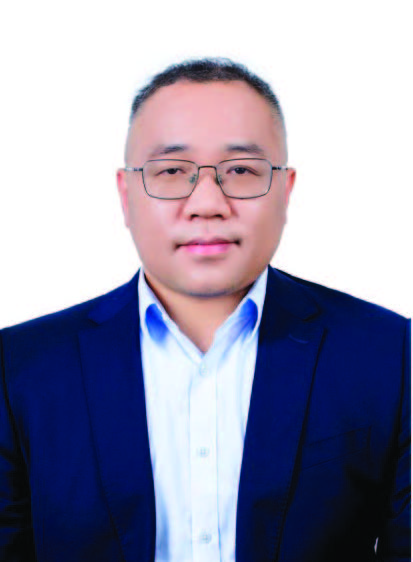}}] 
	{Guanding Yu} (Senior Member, IEEE) received the B.E. and Ph.D. degrees in communication engineering from Zhejiang University, Hangzhou, China, in 2001 and 2006, respectively. He joined Zhejiang University in 2006, and is now a Professor with the College of Information and Electronic Engineering. From 2013 to 2015, he was also a Visiting Professor at the School of Electrical and Computer Engineering, Georgia Institute of Technology, Atlanta, GA, USA. His research interests include integrated sensing and communications (ISAC), mobile edge computing/learning, and machine learning for wireless networks.
	
	Dr. Yu has served as a guest editor of IEEE Communications Magazine special issue on Full-Duplex Communications, an Editor of IEEE Journal on Selected Areas in Communications Series on Green Communications and Networking, and Series on Machine Learning in Communications and Networks, an Editor of IEEE Wireless Communications Letters, a lead Guest Editor of IEEE Wireless Communications Magazine special issue on LTE in Unlicensed Spectrum, an Editor of IEEE Transactions on Green Communications and Networking, and an Editor of IEEE Access. He is now serving as an editor of \emph{IEEE Transactions on Machine Learning in Communications and Networking}. He received the 2016 IEEE ComSoc Asia-Pacific Outstanding Young Researcher Award. He regularly sits on the technical program committee (TPC) boards of prominent IEEE conferences such as ICC, GLOBECOM, and VTC. He also serves as a Symposium Co-Chair for IEEE GlobeCom 2019 and a Track Chair for IEEE VTC 2019'Fall.
\end{IEEEbiography}

\end{document}